\documentclass[
reprint,
groupedaddress,
nofootinbib,
 amsmath,amssymb,
 aps,
prb,
]{revtex4-2}

\usepackage{dcolumn}
\usepackage{graphicx}
\usepackage{hyperref}
\usepackage{color}
\usepackage{tikz}
\usepackage[compat=1.1.0]{tikz-feynhand}
\usepackage{ascmac,amsmath,amsfonts,amsthm,amssymb}
\usepackage{url}
\usepackage[whole]{bxcjkjatype}

\newenvironment{itmbx}[1]{\begin{itembox}[l]{#1}}{\end{itembox}}

\newcommand{\ket}[1]{| #1 \rangle}
\newcommand{\bra}[1]{\langle #1 |}

\newcommand{\fraction}[1]{ \mathrm{frac} \left( #1 \right)}

\newcommand{\MY}[1]{{\color{purple} #1}}

\begin{document}

\preprint{APS/123-QED}

\title{Brillouin zone folding method for quasiperiodic superconductivity in multilayer systems: application to electronic structure and optical responses}

\author{Mao~Yoshii}
\email{mao@g.ecc.u-tokyo.ac.jp}
\affiliation{Department of Applied Physics, The University of Tokyo, Hongo, Tokyo, 113-8656, Japan}

\author{Sota Kitamura}
\affiliation{Department of Applied Physics, The University of Tokyo, Hongo, Tokyo, 113-8656, Japan}

\author{Takahiro Morimoto}
\affiliation{Department of Applied Physics, The University of Tokyo, Hongo, Tokyo, 113-8656, Japan}

\begin{abstract}
    We construct an efficient momentum space approach to the superconductivity in quasiperiodic multilayer systems. To this end, we extend the Brillouin zone (BZ) folding method to the superconducting (SC) phases by formulating the gap equation in the momentum space representation with the BZ folding. We show that the physical observables in quasiperiodic multilayers are generally given by so-called quasiperiodic functions. Consequently, there appear SC order parameters with finite momenta in the BZ folding method, which corresponds to the spatial fluctuation of the SC order in quasiperiodic multilayers.
    We also find a systematic way to compute the physical observables in the BZ folding method with a proper normalization condition using the continued fraction approximation.
    We apply the BZ folding method to a one-dimensional toy model of quasiperiodic superconductors and demonstrate its numerical efficiency compared to the conventional method based on the real space representation with a large system size.
    We also study a quasiperiodic bilayer system of Rice-Mele model and an s-wave superconductor to demonstrate an efficient computation of optical responses of a quasiperiodic superconductor with inversion symmetry breaking.
\end{abstract}

\maketitle

\section{Introduction}\label{sec : Introduction}
Two-dimensional materials have been attracting keen attention because of their variety and high controllability, 
including graphene\cite{doi:10.1126/science.1102896,RevModPhys.81.109,McCann2013,ROZHKOV20161}, transition metal dichalcogenides (TMD) \cite{Manzeli2017} and black phosphorus \cite{Xia2019,doi:10.1073/pnas.1416581112}.
These two-dimensional materials also exhibit various quantum phases such as superconductivity \cite{Xi2015,Qiu2021} and magnetism \cite{Gibertini2019,https://doi.org/10.1002/adfm.201202502}.
In particular, van der Waals (vdW) heterostructures obtained from stacking two-dimensional materials allow for substantial control of their material properties according to the choice of the materials and the way they are stacked \cite{geim2013van,doi:10.1126/science.aac9439}, which includes twisted bilayer graphene \cite{Bistritzer12233,cao2018correlated,PhysRevB.99.165430} and interfaces of transition metal dichalcogenides \cite{wang2020correlated,Akamatsu68}. Such interfaces of two-dimensional materials generally become quasiperiodic systems due to a lattice mismatch or a twist in stacking.

The quasiperiodic systems exhibit interesting superconducting (SC) states \cite{cao2018unconventional,Devarakonda2021,Kezilebieke2020,Li2023,Zhang2023,Uri2023}. The representative example in stacked two-dimensional systems is the SC states observed in twisted bilayer graphene \cite{cao2018unconventional} and trilayer graphene \cite{Uri2023}. 
A twisted bilayer $\mathrm{WSe}_2$ also shows correlated electronic phases, which may lead to SC states \cite{Wang2020}.
Furthermore, quasicrystalline systems are reported to host SC states in the three-dimensional $ \mathrm{Al_{14.9}Mg_{44.1}Zn_{41.0}}$ alloy \cite{kamiya2018discovery} and the two-dimensional $\mathrm{Ta_{1.6}Te}$ flake\cite{Tokumoto2024}. There have been theoretical efforts to study these SC states in both multilayer \cite{LEDWITH2021168646,PhysRevB.103.205413,PhysRevB.103.205414,PhysRevB.103.205415,PhysRevB.103.205416} and quasicrystalline systems \cite{TAKEMORI2023461,PhysRevLett.95.177007,PhysRevB.74.024522,PhysRevB.82.184512}.

Meanwhile, theoretical analysis of multilayer thin films (MLTFs) generally has a difficulty due to the lack of periodicity, where Bloch theorem and the band picture in the momentum space representation cannot be directly applied. One standard approach to these multilayer systems is to extract a finite but sufficiently large region from the quasiperiodic system in the real space representation. While such real space approach is in principle applicable to general quasiperiodic systems, the numerical cost is usually large since one needs to treat a large system size for extrapolation to infinite system size\cite{PhysRevResearch.5.043164,PhysRevResearch.1.022002,PhysRevB.95.024509,doi:10.7566/JPSJ.84.023701,PhysRevB.105.104410,matsubara2023ferromagnetically,PhysRevB.102.115108,PhysRevResearch.5.043164,PhysRevResearch.1.022002,PhysRevB.95.024509,doi:10.7566/JPSJ.84.023701,doi:10.7566/JPSJ.89.074703,PhysRevB.106.064506}.

While several approaches from the reciprocal space have been investigated \cite{K-Niizeki_1990,Niizeki1990,PhysRevResearch.4.013226}, one of the most efficient approaches for the multilayer systems is the Brillouin zone (BZ) folding method \cite{Koshino_2015}. This approach makes use of periodicity for each layer that composes quasiperiodic multilayer systems. Since each layer is periodic, we can define Bloch bases in each layer, whereas the quasiperiodicity is introduced by the interlayer couplings. 
In the BZ folding method, this quasiperiodicity is incorporated as a shift of the momentum between different layers. This treatment based on the momentum space picture allows us to substantially reduce the numerical cost, compared to the conventional real space approach \cite{PhysRevB.99.245401,PhysRevLett.127.147203,PhysRevB.99.165430}. Specifically, one needs to treat a large Hamiltonian with its dimension proportional to the system size in the real space approach, whereas the size of the Hamiltonian is reduced to the number of BZ folding times the size of the Bloch Hamiltonian defined on each layer in the BZ folding method.
Thus the BZ folding method provides a standard tool for studying the electronic structure of multilayer systems with its applicability to general quasiperiodic multilayers and reduced numerical cost.

\begin{figure}
    \centering
    \includegraphics[width = \linewidth]{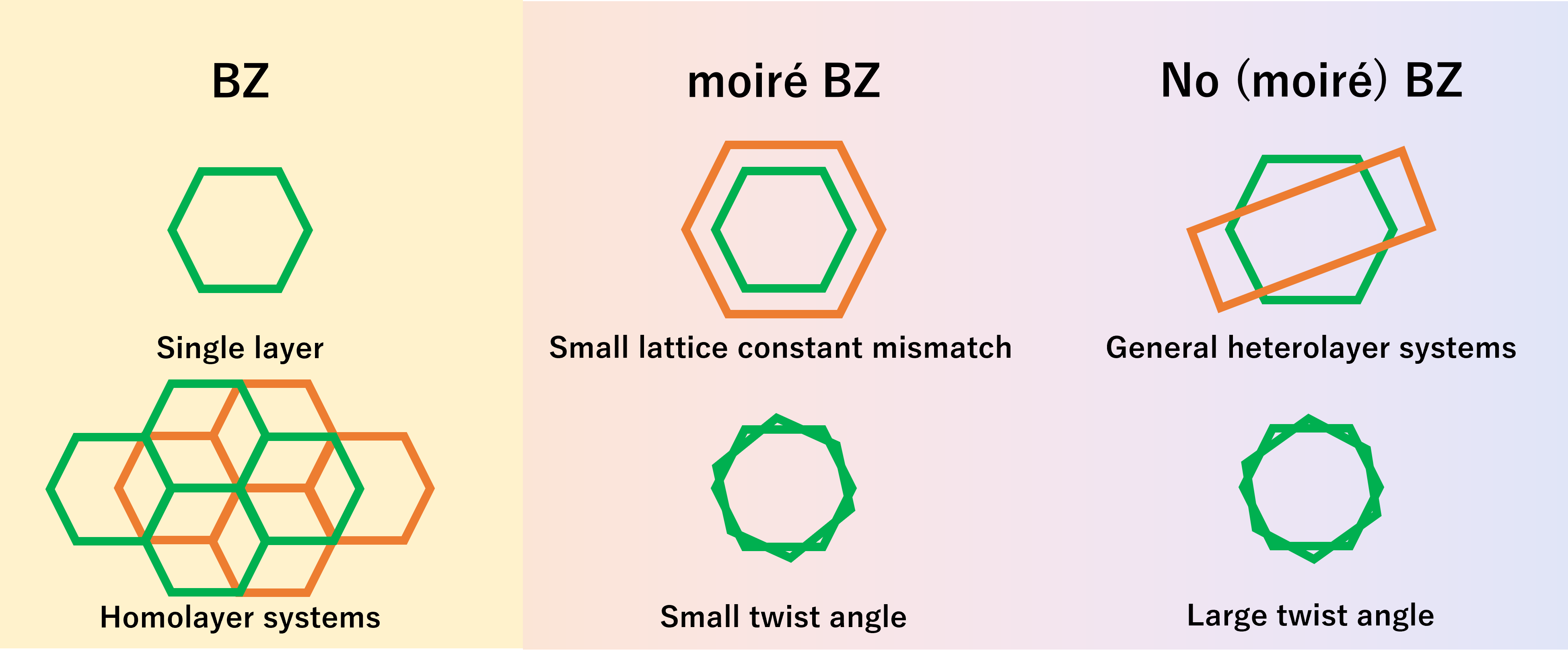}
    \caption{
    Overview of the SC phase in the multilayer thin film systems.
    We consider three classes of multilayer systems. The first class (left) is the systems that are periodic and allow the usual BZ, which includes single layer materials \cite{Xi2015,Ge2014,Zhang2014,Hsu2017,PhysRevX.10.041003} and homolayer systems \cite{doi:10.1126/sciadv.1602390,PhysRevB.107.L161106,PhysRevLett.130.146001,doi:10.1126/science.abm8386,doi:10.7566/JPSJ.83.061014,saito2016superconductivity,Shi2015}. 
    The second class (middle) is the multilayer systems with small quasiperiodicity which allows an efficient description with effective mini (moir\'e) BZs. 
    This is the case when differences between layers such as the twist angle and the lattice constant mismatch are small enough, which includes the magic angle graphene \cite{LEDWITH2021168646,PhysRevB.103.205413,PhysRevB.103.205414,PhysRevB.103.205415,PhysRevB.103.205416}.
    The third class (right) is the systems with large quasiperiodicity where we cannot define the moir\'e BZ. 
    Such systems include homolayers with large twist angles and general heterolayers with large lattice constant mismatches.
    }
    \label{fig : overview_MLTFSC}
\end{figure}

Given the above advantage of the BZ folding method for quasiperiodic multilayer systems, it is also desirable to apply the BZ folding method to the SC states in the multilayer systems, although such application is still limited so far (Fig.~\ref{fig : overview_MLTFSC}). One successful example of such application is the case when we can define the moir\'e BZ. Namely, we can define an effective mini BZ (the moir\'e BZ) when differences between layers such as twist angle and lattice constant are small enough. Such cases include twisted bilayer graphene (and general twisted homolayers) with a small twist angle or some heterolayers of small lattice constant mismatches \cite{Bistritzer12233,LEDWITH2021168646,PhysRevB.103.205413,PhysRevB.103.205414,PhysRevB.103.205415,PhysRevB.103.205416}. In contrast, for more general cases, i.e., multilayer systems with large twist angles and general heterolayer structures with large lattice constant mismatches, one cannot use such a method based on the moir\'e BZ.

In this paper, we extend the BZ folding method to the SC phase in quasiperiodic multilayer systems. By formulating the gap equation in the momentum space representation with the BZ folding, we find that the SC order parameters generally possess a finite center of mass momentum, which corresponds to the spatial fluctuation of the SC order parameter. 
Such SC order with finite momenta arises from the fact that physical observables in quasiperiodic systems are generally given by so-called quasiperiodic functions. 
We also find a systematic way to compute the physical observables in the BZ folding method with a proper normalization condition using the continued fraction approximation. 
We apply the BZ folding method to two examples of one-dimensional bilayer models. 
We first study a model of incommensurate bilayer of SCs and demonstrate the numerical efficiency to capture the SC nature in the quasiperiodic systems. In particular, we demonstrate that the overall behavior of the SC order parameter is efficiently captured with just a few times of the BZ folding. 
We then study a quasiperiodic bilayer system of Rice-Mele model and an s-wave superconductor to show that breaking the inversion symmetry in the s-wave SC leads to the nonlinear optical response called shift current.

This paper is organized as follows.
In Sec.~\ref{sec : Formalism}, we begin with reviewing the BZ folding method for the normal state and then we extend it to the SC state.
In Sec.~\ref{sec : Numerical calculation}, we apply our method to a one-dimensional model of quasiperiodic SCs and show the results of the numerical calculations and study effects of the approximations.
In Sec.~\ref{sec : Discussions}, we give brief discussions. 

\section{Formalism}\label{sec : Formalism}
In this section, we describe our formalism that extends the BZ folding method to the SC phase. For simplicity, we focus on the bilayer system here. We note that it is straightforward to  generalize our method to multilayer systems.

A general form of the noninteracting part of the Hamiltonian in the bilayer system can be written as
\begin{align}\label{eq : original bilayer ham}
    \hat{H} = \hat{H}_1 + \hat{H}_2 + \hat{V} + \hat{V}^\dagger.
\end{align}
Here, $\hat{H}_l$ is a non-interacting Hamiltonian of layer $l$, and they are coupled through the interlayer coupling $\hat{V}$.
We consider a bilayer of superconductors with interlayer coupling and related BZ folding.
Also, we approximate that $\hat{V}$ depends only on the relative distance between two sites as
\begin{align}\label{eq : def interlayer coupling real}
    \hat{V} =& -\sum_{\boldsymbol{r}_1,\boldsymbol{r}_2,\sigma_1\sigma_2} V_{\sigma_1\sigma_2}(\boldsymbol{r}_1-\boldsymbol{r}_2) 
    \hat{c}_{1,\boldsymbol{r}_1,\sigma_1}^\dagger
    \hat{c}_{2,\boldsymbol{r}_2,\sigma_2}.
\end{align}
Here, 
$\boldsymbol{r}_l$ is the position of sites in layer $l$, and
$\sigma_l$ denotes the spin on layer $l$.
While we do not explicitly consider the sublattice degrees of freedom, they can be effectively incorporated as additional layers.

\subsection{Review of the Brillouin zone folding method for the non-interacting systems}

\begin{figure*}
    \centering
    \includegraphics[width = \linewidth]{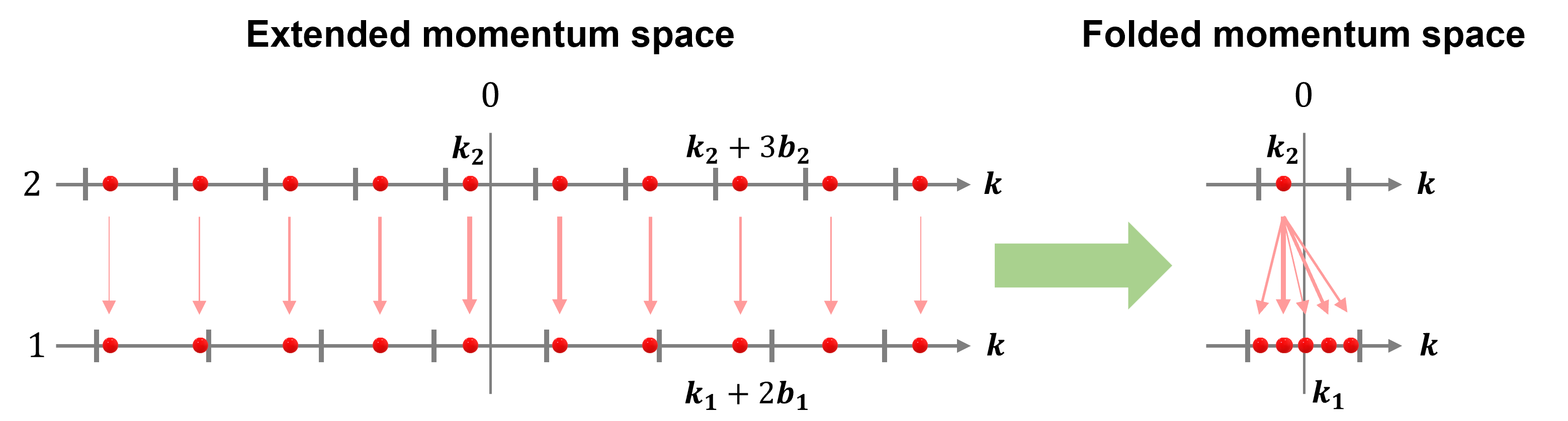}
    \caption{
        Schematic illustration of the interlayer hopping from layer $2$ to layer $1$ in a one-dimensional bilayer system. 
        In the extended momentum space picture (left), we consider the state with the crystal momentum $\boldsymbol{k}_l$ and the reciprocal vector $\boldsymbol{b}_l ( = \boldsymbol{b}_{l,1})$ on each layer.  
        In this picture, the same momenta in the two layers are connected with the interlayer hopping [c.f.~Eq~(\ref{eq : interlayer coupling momentum})]. 
        The interlayer hopping strength $v_{\sigma_1\sigma_2,\boldsymbol{q}}$ rapidly diminishes for larger $|\boldsymbol{q}|$, which is depicted by the width of arrows connecting the two layers. 
        In the BZ folding method (right), we fold back the momenta into the BZ of each layer, and consider only the crystal momentum within the BZ on each layer.
        In this picture, an electron with crystal momentum $\boldsymbol{k}_2$ can hop to multiple $k$ points in the BZ on layer $1$, indicating the breaking of the crystal momentum conservation.
        }
    \label{fig : BZfolding}
\end{figure*}

Before treating SC phases, we first describe the BZ folding method in the normal state \cite{Koshino_2015}.
The Hamiltonian consists of single layer Hamiltonians and the interlayer coupling.
Since single-layer Hamiltonians are described by the momentum space description with the translational symmetry in a straightforward way, we describe how to model the interlayer term in the momentum space in the form of 
\begin{align}
    \hat{V} = -\sum_{\boldsymbol{k}_1,\boldsymbol{k}_2,\sigma_1,\sigma_2} 
    \tilde{V}_{\sigma_1\sigma_2}(\boldsymbol{k}_1,\boldsymbol{k}_2)
    \hat{c}_{1,\boldsymbol{k}_1,\sigma_1}^\dagger
    \hat{c}_{2,\boldsymbol{k}_2,\sigma_2}.
\end{align}
To this end, we use the Bloch bases of each layer,
\begin{align}\label{eq : creation and annihilation op in momentum sp}
    \hat{c}_{l,\boldsymbol{r}_{l},\sigma_l}^\dagger =& \frac{1}{\sqrt{N_l}}\sum_{\boldsymbol{k}_l\in \mathrm{BZ}_l} e^{-i\boldsymbol{k}_l\cdot\boldsymbol{r}_{l}}
    \hat{c}_{l,\boldsymbol{k}_l,\sigma_l}^\dagger,
    \\
    \hat{c}_{l,\boldsymbol{k}_l,\sigma_l}^\dagger =& \frac{1}{\sqrt{N_l}}\sum_{\boldsymbol{r}_l} e^{i\boldsymbol{k}_l\cdot \boldsymbol{r}_{l}}
    \hat{c}_{l,\boldsymbol{r}_{l},\sigma_l}^\dagger.
\end{align}
Here, $k$-summation is taken over the BZ of the layer $l$ $(\mathrm{BZ}_l)$ and $N_l$ is a number of sites on the layer $l$.
Throughout this paper, we assume that $N_l$ is sufficiently large as we can approximate it to be infinite.
The lattice vector $\boldsymbol{r}_l$ can be expressed as
\begin{align}
    \boldsymbol{r}_l = \sum_{\alpha = 1}^{D} n_{l,\alpha} \boldsymbol{a}_{l,\alpha} + \boldsymbol{\tau}_l,
\end{align}
where $D$ is the dimension of the layers and $\boldsymbol{a}_{l,\alpha}$ is the $\alpha$th primitive vector of layer $l$, and $n_{l,\alpha}$ is the integer coefficient.
$\boldsymbol{\tau}_l$ expresses the relative position of layer $l$. 
Inserting this expansion to Eq.~(\ref{eq : def interlayer coupling real}), the interlayer coupling in the momentum space representation is expressed by the real space coupling as
\begin{align}
    \hat{V} =& -\frac{1}{\sqrt{N_1N_2}}
    \sum_{\boldsymbol{r}_1,\boldsymbol{r}_2,\sigma_1,\sigma_2} V_{\sigma_1\sigma_2}(\boldsymbol{r}_1-\boldsymbol{r}_2) 
    \nonumber \\
    & \times \sum_{\boldsymbol{k}_1,\boldsymbol{k}_2} 
    e^{-i\boldsymbol{k}_1\cdot\boldsymbol{r}_{1}+i\boldsymbol{k}_2\cdot\boldsymbol{r}_{2}}
    \hat{c}_{\boldsymbol{k}_1,\sigma_1}^\dagger
    \hat{c}_{\boldsymbol{k}_2,\sigma_2}.
\end{align}
Assuming that the system is incommensurate (or moir\'e superlattice is large enough to be regarded as an incommensurate structure), we define the inverse Fourier transformation of $V_{\sigma\sigma^\prime}$ as
\begin{align}
    V_{\sigma\sigma^\prime}\left(\boldsymbol{r}\right)
    =&
    \frac{ \sqrt{S_1S_2}}{(2\pi)^D}
    \int d^Dq
    v_{\sigma\sigma^\prime,\boldsymbol{q}}e^{i\boldsymbol{q}\boldsymbol{r}} , \label{eq : Fourier V1}
    \\
    v_{\sigma\sigma^\prime,\boldsymbol{q}}
    =&
    \frac{1}{\sqrt{S_{1}S_{2}}}
    \int d^Dr
    V_{\sigma\sigma^\prime}\left(\boldsymbol{r}\right) e^{-i\boldsymbol{q}\boldsymbol{r}}. \label{eq : Fourier V2}
\end{align}
Here, $S_l$ is the volume of the unit cell on layer $l$.
Using this expansion,
we obtain
\begin{align}
    \hat{V} 
    =&-
    \frac{1}{(2\pi)^D}\sqrt{\frac{S_1S_2}{N_1N_2}}
    \sum_{\boldsymbol{r}_1,\boldsymbol{r}_2,\sigma_1,\sigma_2} \int d^Dq 
    v_{\sigma_1\sigma_2,\boldsymbol{q}} 
    e^{i\boldsymbol{q}\cdot(\boldsymbol{r}_{1}-\boldsymbol{r}_{2})}
    \nonumber
    \\
    & \times \sum_{\boldsymbol{k}_1,\boldsymbol{k}_2} 
    e^{-i\boldsymbol{k}_1\cdot\boldsymbol{r}_{1}+i\boldsymbol{k}_2\cdot\boldsymbol{r}_{2}}
    \hat{c}_{1,\boldsymbol{k}_1,\sigma_1}^\dagger
    \hat{c}_{2,\boldsymbol{k}_2,\sigma_2}
    \nonumber\\
    =&-
    \sum_{\boldsymbol{k}_1,\boldsymbol{k}_2,\sigma_1,\sigma_2}
    \tilde{V}_{\sigma_1\sigma_2}(\boldsymbol{k}_1,\boldsymbol{k}_2)
    \hat{c}_{1,\boldsymbol{k}_1,\sigma_1}^\dagger
    \hat{c}_{2,\boldsymbol{k}_2,\sigma_2},
\end{align}
where $\tilde{V}_{\sigma_1\sigma_2}(\boldsymbol{k}_1,\boldsymbol{k}_2)$ is defined as
\begin{align}\label{eq : interlayer coupling momentum}
    \tilde{V}_{\sigma_1\sigma_2}(\boldsymbol{k}_1,\boldsymbol{k}_2)
    =&
    \sum_{\boldsymbol{G}_l \in \mathcal{B}_l }
 v_{\sigma_1\sigma_2,\boldsymbol{k}_1+\boldsymbol{G}_{1}}
    \nonumber
    \\ & \times
    \delta_{
    \boldsymbol{k}_1+\boldsymbol{G}_1,
    \boldsymbol{k}_2+\boldsymbol{G}_{2}}
    e^{i\boldsymbol{G}_1\cdot\boldsymbol{\tau}_{1}-i\boldsymbol{G}_2\cdot\boldsymbol{\tau}_{2}}.
\end{align}
See Appendix \ref{sec : Interlayer coupling} for detailed derivation.
Here, $\mathcal{B}_l$ is a set of reciprocal vectors on layer $l$ defined with the primitive reciprocal vectors $\boldsymbol{b}_{l,\alpha}$ of $\boldsymbol{a}_{l,\alpha}$ that satisfy $\boldsymbol{a}_{l,\alpha}\cdot \boldsymbol{b}_{l,\beta} = 2\pi\delta_{\alpha,\beta}$ as $ \mathcal{B}_l = \{ \sum_{\alpha = 1}^{D} m_{l,\alpha} \boldsymbol{b}_{l,\alpha}  | m_{l,\alpha} \in \mathbb{Z} \}$.
In the above equation, the momentum is shifted as $\boldsymbol{k}_1 \to \boldsymbol{k}_1 +\boldsymbol{G}_1$, which we call the BZ folding.
Using fermion operators in different BZs,
\begin{align}\label{eq : quasiperiodicity of c}
    \hat{c}_{l,\boldsymbol{k}_l+\boldsymbol{b}_{l,\alpha},\sigma_l} =
    e^{-i\boldsymbol{b}_{l,\alpha}\cdot \boldsymbol{\tau}_l}\hat{c}_{l,\boldsymbol{k}_l,\sigma_l},
\end{align}
we can eliminate the phase factor in Eq.~\eqref{eq : interlayer coupling momentum} as
\begin{align}\label{eq : Interlayer coupling extended}
    \hat{V} 
    =&-
    \sum_{\boldsymbol{k}_1,\boldsymbol{k}_2,\sigma_1,\sigma_2}
    \sum_{\boldsymbol{G}_1,\boldsymbol{G}_2}
    v_{\sigma_1\sigma_2,\boldsymbol{k}_1+\boldsymbol{G}_1}
    \nonumber\\
     & \times
     \delta_{
    \boldsymbol{k}_1+\boldsymbol{G}_1,
    \boldsymbol{k}_2+\boldsymbol{G}_2}
    \hat{c}_{1,\boldsymbol{k}_1+\boldsymbol{G}_1,\sigma_1}^\dagger
    \hat{c}_{2,\boldsymbol{k}_2+\boldsymbol{G}_2,\sigma_2}.
\end{align}
The above expression includes the fermion operators outside the first BZ, and corresponds to the extended momentum space picture as illustrated in the left panel in Fig.~\ref{fig : BZfolding}.
In the extended momentum space picture, the states with the same momenta in the two layers are connected with the interlayer hopping [Eq.~(\ref{eq : interlayer coupling momentum})]. 
In contrast, in the BZ folding method, we fold back the momenta into the BZ of each layer and consider only the crystal momentum within the BZ on each layer (right panel in Fig.~\ref{fig : BZfolding}).
In this picture, an electron with crystal momentum $\boldsymbol{k}_2$ can hop to multiple k-points in the BZ for the other layer, indicating the breaking of the crystal momentum conservation.
Using the interlayer coupling modeled above and $\hat{H}_l = 
\sum_{\boldsymbol{k}_l \in \mathrm{BZ}_l} 
\sum_{\sigma_l,\sigma_l^\prime}
\hat{c}_{l,\boldsymbol{k}_l,\sigma_l}^\dagger h_{l,\boldsymbol{k}_l,\sigma_l\sigma_l^\prime}\hat{c}_{l,\boldsymbol{k}_l,\sigma_l^\prime}$, we can also reconstruct the Hamiltonian in Eq.~(\ref{eq : original bilayer ham}) as
\begin{align}\label{eq : Ham_momentum}
\hat{H}_{\sigma\sigma^\prime}
    =& 
    \begin{pmatrix}
    \hat{\boldsymbol{c}}_{1,\sigma}^\dagger
        &
\hat{\boldsymbol{c}}_{2,\sigma}^\dagger
    \end{pmatrix}
    \begin{pmatrix}
        H_{1,\sigma\sigma^\prime} & \tilde{V}_{\sigma\sigma^\prime} \\
        \tilde{V}_{\sigma\sigma^\prime}^\dagger & H_{2,\sigma\sigma^\prime}
    \end{pmatrix}
    \begin{pmatrix}
        \hat{\boldsymbol{c}}_{1,\sigma^\prime}
        \\
        \hat{\boldsymbol{c}}_{2,\sigma^\prime}
    \end{pmatrix}
    ,
    \\
    \hat{\boldsymbol{c}}_{l,\sigma}^\dagger = &(\ldots, \hat{c}_{l,\boldsymbol{k},\sigma}^\dagger,\ldots)
    \quad
    (\boldsymbol{k}_l \in \mathrm{BZ}_l ).
\end{align}
Here, $H_{l,\sigma_l,\sigma_l^\prime}$ are the collections of the single layer Hamiltonians defined as
\begin{align}
    H_{l,\sigma_l\sigma_l^\prime}=& 
\begin{pmatrix}
        \ddots & & \\
         & h_{l,\boldsymbol{k}_l,\sigma_l\sigma_l^\prime} & \\
          & & \ddots 
    \end{pmatrix}
    \quad (\boldsymbol{k}_l \in \mathrm{BZ}_l).
\end{align}
These two blocks are connected with the interlayer coupling $\tilde{V}_{\sigma_1\sigma_2}$ of the form
\begin{align}
    \tilde{V}_{\sigma_1\sigma_2}
    =&
    \begin{pmatrix}
         \ddots & & & & \\
         & \tilde{V}_{\sigma_1\sigma_2}(\boldsymbol{k}_1,\boldsymbol{k}_2) & \cdots & \tilde{V}_{\sigma_1\sigma_2}(\boldsymbol{k}_1,\boldsymbol{k}_2^\prime) & \\
         & \vdots & \ddots & \vdots & \\
         & \tilde{V}_{\sigma_1\sigma_2}(\boldsymbol{k}_1^\prime,\boldsymbol{k}_2) & \cdots & \tilde{V}_{\sigma_1\sigma_2}(\boldsymbol{k}_1^\prime,\boldsymbol{k}_2^\prime) & \\
         & & & & \ddots 
    \end{pmatrix},
\end{align}
where $\boldsymbol{k}_l \in \mathrm{BZ}_l$.
In this form, all $\boldsymbol{k}_1$ and $\boldsymbol{k}_2$ are connected with each other, however not all $\tilde{V}_{\sigma\sigma^\prime}(\boldsymbol{k}_1,\boldsymbol{k}_2)$ take nonzero value as indicated by the Kronecker's delta in Eq.~(\ref{eq : interlayer coupling momentum}).
Since $\tilde{V}_{\sigma\sigma^\prime}(\boldsymbol{k}_1,\boldsymbol{k}_2)$ is finite only when $\boldsymbol{k}_1 + \boldsymbol{G}_1 = \boldsymbol{k}_2 + \boldsymbol{G}_2$, we can block diagonalize the Hamiltonian $\hat{H}$ by selecting one $\boldsymbol{k}$ and momenta connected by the reciprocal vectors to make
\begin{align}\label{eq : Infinite dimensional Hamiltonian in momentum sp}
    H_{\sigma\sigma^\prime}(\boldsymbol{k}) =& 
    \begin{pmatrix}
        H_{1,\mathcal{B}_2,\sigma\sigma^\prime}(\boldsymbol{k}) & V_{\mathcal{B}_2,\mathcal{B}_1,\sigma\sigma^\prime}(\boldsymbol{k}) \\
        V_{\mathcal{B}_1,\mathcal{B}_2,\sigma\sigma^\prime}(\boldsymbol{k}) & H_{2,\mathcal{B}_1,\sigma\sigma^\prime}(\boldsymbol{k})
    \end{pmatrix}.
\end{align}
Here, diagonal block is defined as
\begin{align}
    H_{l,\mathcal{B}_{l^\prime},\sigma\sigma^\prime}(\boldsymbol{k})=& 
    \begin{pmatrix}
        \ddots & & \\
         & h_{l,\boldsymbol{k}+\boldsymbol{G}_{l^\prime},\sigma\sigma^\prime} & \\
          & & \ddots 
    \end{pmatrix}
    ,
\end{align}
where $\boldsymbol{G}_{l^\prime} \in \mathcal{B}_{l^\prime}$.
These diagonal blocks are connected with the interlayer coupling $V_{\mathcal{B}_2,\mathcal{B}_1,\sigma\sigma^\prime}$ that has the form,
\begin{align}
    V_{\mathcal{B}_2,\mathcal{B}_1, \sigma_1,\sigma_2}(\boldsymbol{k})
    =&
    \begin{pmatrix}
          & \vdots & \\
         \cdots & v_{\sigma_1\sigma_2,\boldsymbol{k}+\boldsymbol{G}_1 + \boldsymbol{G}_2} & \cdots \\
         & \vdots & 
    \end{pmatrix}.
\end{align}
The complete description of the system can be obtained from the eigenspectrum of $H_{\sigma\sigma^\prime}(\boldsymbol{k})$ for all k-points, while this is usually difficult since the dimension of $\hat{H}$ is infinite.
In practical calculations, we approximate this matrix by selecting subset $\boldsymbol{B}_l = \{ \tilde{\boldsymbol{b}}_{l,1},\ldots, \tilde{\boldsymbol{b}}_{l,M_l} \}$ of $\mathcal{B}_l$ to make following Hamiltonian,
\begin{align}\label{eq : Moire Ham normal}
    H_{\boldsymbol{B}_2,\boldsymbol{B}_1,\sigma\sigma^\prime}(\boldsymbol{k}) =& \begin{pmatrix}
        H_{1,\boldsymbol{B}_2,\sigma\sigma^\prime}(\boldsymbol{k}) & V_{\boldsymbol{B}_1,\boldsymbol{B}_2,\sigma\sigma^\prime}(\boldsymbol{k}) \\
        V_{\boldsymbol{B}_1,\boldsymbol{B}_2,\sigma\sigma^\prime}(\boldsymbol{k})^\dagger & H_{2,\boldsymbol{B}_1,\sigma\sigma^\prime}(\boldsymbol{k})
    \end{pmatrix}.
\end{align}
To obtain physical observables approximately well, we consider a set of Hamiltonians with limited number of the reference $k$-points as
\begin{align}\label{eq : folded Hamiltonian}
    H_{eff}&=\bigoplus_{i=1}^{I}
    H_{\boldsymbol{B}_{i,2},\boldsymbol{B}_{i,1},\sigma\sigma^\prime}(\boldsymbol{k}_i),
\end{align}
where we sample $I$ points $\{ \boldsymbol{k}_1, \ldots \boldsymbol{k}_I \}$ from the k-space. 
$H_{eff}$ gives a submatrix of the Hamiltonian matrix for the original Hamiltonian $\hat{H}_{\sigma\sigma^\prime}$ in Eq.~\eqref{eq : Ham_momentum}.
In the BZ folding method, we call $H_{\boldsymbol{B}_2,\boldsymbol{B}_1,\sigma\sigma^\prime}(\boldsymbol{k})$ the folded Hamiltonian.
Since $\boldsymbol{k}+\boldsymbol{G}_2$ and $\boldsymbol{k}+\boldsymbol{G}_1$ are dense in BZ$_1$ and BZ$_2$, 
we can obtain physical observables with $H_{eff}$ which well approximate those with $H_{\sigma\sigma^\prime}$ as far as $I$ and $|\boldsymbol{B}_i|$ are sufficiently large. 
For good approximation, we need to select $\boldsymbol{B}_{i,l}$ from $\mathcal{B}_l$ appropriately for given reference points, and $(\boldsymbol{k}_i + \boldsymbol{B}_{i,l}) \cap (\boldsymbol{k}_j + \boldsymbol{B}_{j,l}) = \varnothing$ should hold for any $i,j \in I$.
Also, we need to be careful about the normalization condition for physical observables since we essentially take a subset of the basis sets for $H_{eff}$; a systematic way to achieve a suitable normalization condition is discussed in Sec.~\ref{subsec: normalization}.
We note that the reason why the Hamiltonian in Eq.~\eqref{eq : folded Hamiltonian} well approximates $H_{\sigma\sigma^\prime}(\boldsymbol{k})$ by suitably selecting a finite number of foldings $\boldsymbol{B}_i$ relies on the nature of the interlayer coupling $v(\boldsymbol{q})$ for MLTFs that decays rapidly against $|\boldsymbol{q}|$, which we will explain in Appendix~\ref{sec : decay of V}.
To summarize, the BZ folding method can be applied to general quasiperiodic multilayer systems with reduced numerical costs by making use of the periodicity of each layer, 
and provides an efficient tool to study 
the electronic structure and physical properties of general quasiperiodic multilayer systems. 
We note that this method cannot be applied to general quasicrystals because we cannot decompose the Hamiltonian of general quasicrystals into periodic intralayer terms and interlayer coupling terms.

\subsection{Brillouin zone folding for interacting systems}
In this subsection, we consider the folded Hamiltonian for interacting systems. 
For simplicity, let us introduce the contact interaction on each layer,
\begin{align}
    H_{l,int} = -g_l \sum_{\boldsymbol{r}_l} 
    \hat{c}_{l,\boldsymbol{r}_l,\uparrow}^\dagger
    \hat{c}_{l,\boldsymbol{r}_l,\uparrow}
    \hat{c}_{l,\boldsymbol{r}_l,\downarrow}^\dagger
    \hat{c}_{l,\boldsymbol{r}_l,\downarrow}.
\end{align}
Here, $g_l$ is the amplitude of the contact interaction on layer $l$.
As explained in Appendix \ref{sec : Intralayer interaction}, the interaction on layer $l$ has the following form in the momentum space representation,
\begin{align}\label{eq : contact int qp}
    H_{l,int}
    =& -\frac{g_l}{N_l}
    \sum_{\boldsymbol{q}}
    \sum_{\boldsymbol{k}_1,\boldsymbol{k}_2}
    \hat{c}_{l,\boldsymbol{k}_1+\boldsymbol{q},\uparrow}^\dagger
    \hat{c}_{l,\boldsymbol{k}_2-\boldsymbol{q},\downarrow}^\dagger
    \hat{c}_{l,\boldsymbol{k}_2,\downarrow}
    \hat{c}_{l,\boldsymbol{k}_1,\uparrow},
\end{align}
with system size $N_l$. 
Next, we apply the mean-field approximation to describe the SC phases.
If we turn off the interlayer coupling and focus on layer $l$, the Cooper pairs take a nonzero value only when $\boldsymbol{k}_1+\boldsymbol{k}_2$ is zero in the BZ of layer $l$. 
In the extended momentum space picture, $\boldsymbol{k}_1+\boldsymbol{k}_2 = \boldsymbol{0}$ inside the BZ means $\boldsymbol{k}_1+\boldsymbol{k}_2 \in \mathcal{B}_l$ in the extended momentum space due to the periodicity on each layer.
Also in the incommensurate system, one may expect that the order parameter follows the quasiperiodicity of the system as long as the interaction is weak enough to ignore the spontaneous spatial symmetry breaking effect.
In this limit, SC gap function is a quasiperiodic function as explained in Appendix~\ref{sec : Functions defined on the MLTFs}.
Therefore, in this paper, we utilize the Fourier transformation of the quasiperiodic function explained in Appendix~\ref{sec : Fourier expansion of quasiperiodic functions} to specify possible center of mass momenta in the MLTFs.
In bilayer systems, one can expand the SC gap function as
\begin{align}\label{eq : Fourier of Delta}
    \Delta^{real}_l(\boldsymbol{r}) =&
    -g_l \langle \hat{c}_{l,\boldsymbol{r},\downarrow}\hat{c}_{l,\boldsymbol{r},\uparrow} \rangle
    =
    \sum_{\boldsymbol{G}}
    e^{
    -i\boldsymbol{r}
    \cdot
    \boldsymbol{G}
    }
    \Delta_{l,\boldsymbol{G}},
\end{align}
with $\boldsymbol{G} \in \mathcal{B}_1 + \mathcal{B}_2 $.
This indicates that in bilayer systems (with inversion and time reversal symmetry), the center of mass momentum of the Cooper pairs can generally take a sum of  the reciprocal vectors of two layers as $\boldsymbol{k}_1+\boldsymbol{k}_2 \in \mathcal{B}_1 + \mathcal{B}_2$.
Hence, the intralayer coupling under the mean field approximation is written as
\begin{align}
    H_{l,int}
    =& -\frac{g_l}{N_l}
    \sum_{\boldsymbol{b}_M}
    \sum_{\boldsymbol{k},\boldsymbol{k}^\prime}
    \hat{c}_{l,\boldsymbol{k},\uparrow}^\dagger 
    \hat{c}_{l,\boldsymbol{b}_M-\boldsymbol{k},\downarrow}^\dagger
    \hat{c}_{l,\boldsymbol{b}_M-\boldsymbol{k}^\prime,\downarrow}
    \hat{c}_{l,\boldsymbol{k}^\prime,\uparrow}
    \\
    =& 
    \sum_{\boldsymbol{b}_M}
    \Bigg[
    \sum_{\boldsymbol{k}}
    \hat{c}_{l,\boldsymbol{k},\uparrow}^\dagger 
    \hat{c}_{l,\boldsymbol{b}_M-\boldsymbol{k},\downarrow}^\dagger
    \Delta_{l,\boldsymbol{b}_M}
    \nonumber \\
    &\qquad 
    +
    \sum_{\boldsymbol{k}^\prime}
    \hat{c}_{l,\boldsymbol{b}_M-\boldsymbol{k}^\prime,\downarrow}
    \hat{c}_{l,\boldsymbol{k}^\prime,\uparrow}
    \Delta_{l,\boldsymbol{b}_M}^\dagger
    \Bigg],
\end{align}
where the center of mass momentum of the Copper pair runs over $\boldsymbol{b}_M \in \mathcal{B}_1 + \mathcal{B}_2$.
Here, the SC gap function is defined as
\begin{align}\label{eq : gap function finite momentum}
    \Delta_{l,\boldsymbol{b}_M} = 
    -\frac{g_l}{N_l}
    \sum_{\boldsymbol{k}^\prime}
    \langle
    \hat{c}_{l,\boldsymbol{b}_M-\boldsymbol{k}^\prime,\downarrow}
    \hat{c}_{l,\boldsymbol{k}^\prime,\uparrow}
    \rangle,
\end{align}
with 
\begin{align}
    \langle \cdots \rangle = \frac{\mathrm{Tr}\left\{ \cdots e^{-\beta \hat{H} }\right\}}{\mathrm{Tr}\left\{e^{-\beta \hat{H} }\right\}}.
\end{align}
Here, $\hat{H}$ is the (folded) Hamiltonian in Eq.~(\ref{eq : folded Hamiltonian}).
$\beta$ is the inverse temperature. 
While original infinite dimensional Hamiltonian in Eq.~(\ref{eq : Ham_momentum}) is invariant against the momentum shift $\boldsymbol{k} \mapsto \boldsymbol{k} + \boldsymbol{G} \ (\boldsymbol{G} \in \mathcal{B}_1 + \mathcal{B}_2)$, truncated Hamiltonian in Eq.~(\ref{eq : folded Hamiltonian}) is not.
Indeed, even though $H_{1,\mathcal{B}_2,\sigma\sigma^\prime}(\boldsymbol{k})$ and $H_{1,\mathcal{B}_2,\sigma\sigma^\prime}(\boldsymbol{k}+\boldsymbol{b}_{1,\alpha})$ are equivalent under a unitary transformation which permutes $\mathcal{B}_2 + \boldsymbol{b}_{1,\alpha}$ to $\mathcal{B}_2$,  $H_{1,\boldsymbol{B}_2,\sigma\sigma^\prime}(\boldsymbol{k}) $ and
$ H_{1,\boldsymbol{B}_2,\sigma\sigma^\prime}(\boldsymbol{k}+\boldsymbol{b}_{1,\alpha})$ are not, because not all elements in $\boldsymbol{B}_2 + \boldsymbol{b}_{1,\alpha}$ are included in $\boldsymbol{B}_2$.
Therefore, $e^{-\beta \hat{H}}$ defined by the folded Hamiltonian is not invariant against $\boldsymbol{k} \mapsto \boldsymbol{k} + \boldsymbol{G}$, and also $\langle \hat{c}_{l,\boldsymbol{b}_M-\boldsymbol{k}^\prime,\downarrow}\hat{c}_{l,\boldsymbol{k}^\prime,\uparrow}\rangle$, due to $\langle \cdots \rangle$. 
In principle, this requires that the entire $k$-space should be incorporated in the $k$-summation in Eq.~(\ref{eq : gap function finite momentum}).
In practice, we take a summation over a sufficiently large area in the $k$-space.

\subsection{Folded Hamiltonian for the bilayer superconductors}
Now that we represented an interaction term in the BZ folding picture, in this section, we construct a folded Hamiltonian for the heterobilayer superconductor of this interaction.
When we turn off the interlayer coupling, each folded band becomes independent. 
In this case, we can block diagonalize the Hamiltonian for each folded band.
The SC gap function $\Delta$ of each block becomes independent of each other,
and the corresponding Bogoliubov-de Gennes (BdG) Hamiltonian is given by
\begin{align}
    H_{l,BdG}(\boldsymbol{k}+\boldsymbol{b}) = \begin{pmatrix}
        H_l(\boldsymbol{k}+\boldsymbol{b}) & \Delta \\
        \Delta^\dagger & -H_l(-\boldsymbol{k}-\boldsymbol{b})
    \end{pmatrix}.
\end{align}
Since the folded Hamiltonian in the BdG form must satisfy the particle hole symmetry of superconductors, when we choose folding $\boldsymbol{b}$ in the electron part, we should also choose $-\boldsymbol{b}$ in the hole part.
For this reason, the folded Hamiltonian for the SC phase is given by
\begin{align}
    &\hat{H}_{BdG,\boldsymbol{B}_2,\boldsymbol{B}_1}
    \Psi_{\boldsymbol{B}_2,\boldsymbol{B}_1}^\dagger(\boldsymbol{k})
    H_{BdG,\boldsymbol{B}_2,\boldsymbol{B}_1}(\boldsymbol{k})
    \Psi_{\boldsymbol{B}_2,\boldsymbol{B}_1}(\boldsymbol{k}),
\end{align}
with
\begin{align}\label{eq : moire BdG}
    &
    H_{BdG,\boldsymbol{B}_2,\boldsymbol{B}_1}(\boldsymbol{k})
    \nonumber \\
    &= \begin{pmatrix}
        H_{\boldsymbol{B}_1,\boldsymbol{B}_2,\uparrow,\uparrow}(\boldsymbol{k}) & \Delta \\
        \Delta^\dagger & -H_{-\boldsymbol{B}_1,-\boldsymbol{B}_2,\downarrow,\downarrow}(-\boldsymbol{k})
    \end{pmatrix}
\end{align}
and
\begin{align}\label{eq : Nambu spinor extended momentum sp}
    \Psi_{\boldsymbol{B}_2,\boldsymbol{B}_1}(\boldsymbol{k}) &=
    \begin{pmatrix}
        \hat{\boldsymbol{c}}_{1,\boldsymbol{k},\boldsymbol{B}_2,\uparrow}
        \\
        \hat{\boldsymbol{c}}_{2,\boldsymbol{k},\boldsymbol{B}_1,\uparrow}
        \\
        \hat{\boldsymbol{c}}_{1,-\boldsymbol{k},-\boldsymbol{B}_2,\downarrow}^\dagger
        \\
        \hat{\boldsymbol{c}}_{2,-\boldsymbol{k},-\boldsymbol{B}_1,\downarrow}^\dagger
    \end{pmatrix},
\end{align}
where $ \hat{\boldsymbol{c}}_{l,\boldsymbol{k},\boldsymbol{B}_{l^\prime},\sigma} = {}^t(\hat{c}_{l,\boldsymbol{k} + \tilde{\boldsymbol{b}}_{l^\prime,1},\sigma},\ldots,\hat{c}_{\boldsymbol{k} + \tilde{\boldsymbol{b}}_{l^\prime,M_{l^\prime}},\sigma})$.
Matrix form of SC gap function $\Delta$ is written down as,
\begin{align}\label{eq : order parameter in matrix form}
    \Delta
    =
    \begin{pmatrix}
        \Delta_{1} & O \\
        O & \Delta_{2}
    \end{pmatrix},
\end{align}
where $\Delta_i$ for $i=1,2$ is the gap function matrix for each layer and is given by
\begin{align}
    \Delta_i &=
    \begin{pmatrix}
        \Delta_{i,\boldsymbol{0}} & \Delta_{i,\tilde{\boldsymbol{b}}_{j,1}-\tilde{\boldsymbol{b}}_{j,2}} & \cdots & \cdots & \Delta_{i,\tilde{\boldsymbol{b}}_{j,1}-\tilde{\boldsymbol{b}}_{j,M_j}} 
        \\
        \Delta_{i,\tilde{\boldsymbol{b}}_{j,2}-\tilde{\boldsymbol{b}}_{j,1}} & \ddots & & & \vdots 
        \\
        \vdots & & \ddots & & \vdots 
        \\
        \vdots & & & \ddots & \vdots 
        \\
        \Delta_{i,\tilde{\boldsymbol{b}}_{j,M_j}-\tilde{\boldsymbol{b}}_{j,1}} & \cdots & \cdots & \cdots & \Delta_{i,\boldsymbol{0}} 
        \\
       \end{pmatrix}.
\end{align}
Here, $j$ denotes the layer different from the layer $i$ (in the bilayer case, $(i,j) = (1,2)$ or $(2,1)$).
Solving \eqref{eq : gap function finite momentum} and \eqref{eq : moire BdG} self-consistently, we obtain SC the gap functions.

\subsection{Convergence of BZ folding procedure}\label{sec : Convergence of error}
The present BZ folding method for computing the gap functions in moir\'e heterostructures is an approximation due to two factors as we explain below.

The first approximation arises from the finiteness of the integration area over $\boldsymbol{k}$.
In Eq.~(\ref{eq : gap function finite momentum}), we perform the $k$-summation of the superconducting pairing amplitude over the BZ.
Since we cannot define the BZ in the quasiperiodic systems, it is necessary to perform the $k$-summation over the infinite momentum space, which is usually not tractable. 
In this work, we approximate the $k$ summation in Eq.~(\ref{eq : gap function finite momentum}) over a sufficiently large but finite area in the $k$-space instead of the infinite $k$-space.

The second approximation is a finite number of the BZ folding.
When a bilayer system is purely incommensurate, we need to fold the BZ in infinite times in principle.
We approximate this infinite BZ folding with a finite number of BZ folding.
In practice, the BZ folding just a few times gives a reasonable result for many cases.

\subsection{Normalization for physical quantities \label{subsec: normalization}}
For obtaining physical quantities, we need to find a proper normalization for observables since we no longer have well defined BZ in the present approach.
When the integrated quantity is defined only on one periodic layer, one can normalize the quantity using the BZ of the layer.
On the other hand, observables including interlayer matrix elements  should be defined on the entire multilayer systems, so we have to normalize them paying attention to all layers.
Below we describe a systematic way to obtain a proper normalization condition.

Let us consider a bilayer system of 1D chains of the lattice constant $a_1$ and $a_2$ below.
In the case of $V_0 = 0$, one can use a conventional normalization using the BZ.
Namely, if
the layers are decoupled, we can normalize the calculated quantity independently on each layer.
If the integration area is $[-\pi,\pi]$, this is $a_1$ times larger than $\mathrm{BZ}_1$ and $a_2$ times larger than $\mathrm{BZ}_2$ ($\mathrm{BZ}_i$: BZ for the $i$th layer $[-\pi/a_i,\pi/a_i]$), thus one can normalize the quantity by dividing by $a_1$ and $a_2$.
However, with finite interlayer coupling, one cannot normalize by $a_1$ nor $a_2$ simply because $a_1 \neq a_2$ in general.
To resolve this issue, we need another degree of freedom, that is, the BZ foldings.
With the BZ folding, the dimension of the folded Hamiltonian is effectively enlarged with the number of the folded energy bands.
In commensurate systems, this effect completely offsets the difference of BZs. 
For example, when $(a_1,a_2) = (3,5)$, the bands from the layer $1$ are folded by $a_2 = 5$ times and the bands from the layer $2$ chain are folded by $a_1 = 3$ times.
This also means that the integration area on layer $1$ is effectively multiplied by $a_2$ and the integration area on layer $2$ is effectively multiplied by $a_1$.
As a result, the effective integration area is $a_1a_2$ times larger than $\mathrm{BZ}_l$.
Hence, the integration area on both layers is multiplied by $a_1a_2$ equally, by which we can normalize the physical quantities.

Next, we apply the same argument to the incommensurate system.
Let $f_l$ be the number of foldings on layer $l$. Then the integration area on layer $1$ is $a_1f_1$ times larger than $\mathrm{BZ}_1$ and the integration area on layer $2$ is $a_2f_2$ times larger than $\mathrm{BZ}_2$.
When these multiplicities are the same, $a_1f_1 = a_2f_2$.
Hence, if 
\begin{align}
    \frac{f_2}{f_1} = \frac{a_1}{a_2},
\end{align}
we could normalize the quantity properly. However, we cannot find integers $f_l$ that satisfy this condition because $a_1/a_2$ is irrational. 
Instead, we satisfy this condition approximately using the continued fraction approximation which is a systematic way to approximate irrational numbers with high accuracy. When the irrational number $\gamma = [\gamma_0; \gamma_1,\gamma_2,\ldots]$ is approximated by coprime numbers $p_n$ and $q_n$ as $p_n/q_n = [\gamma_0; \gamma_1,\gamma_2,\ldots, \gamma_n]$, the error is bounded as $|\gamma - p_n/q_n| < 1/q_n^2$.
For example, for $(a_1,a_2) = (6,\sqrt{24})$, we approximate $a_1/a_2$ as
\begin{align}
    \frac{a_1}{a_2} = [1;4,2,4,2,4,2,\ldots]
    \sim
    1+\frac{1}{
    4+\frac{1}{2}
    }
    =
    \frac{11}{9}.
\end{align}
In fact, $(11/9)/(a_1/a_2) = 0.9979402\ldots$ so the difference is $0.2\%$.
This indicates that one can offset the difference of lattice constant approximately well, by folding the layer $1$ by $9$ times and the layer $2$ by $11$ times.

\section{Application to 1D superconductors}\label{sec : Numerical calculation}
In this section, we show the result of the numerical calculations for a model of 1D superconductors.

We consider a one-dimensional system, where the single layer Hamiltonian has the following form
\begin{align}
    \hat{H}_l &= 
    t
    \sum_{\boldsymbol{r}_l}
    \sum_{\sigma}
    \Big[\hat{c}^\dagger_{l,\boldsymbol{r}_l+\boldsymbol{a_l},\sigma}
    \hat{c}_{l,\boldsymbol{r}_l,\sigma}+h.c.\Big]\nonumber\\
    &-
    g_l
    \sum_{\boldsymbol{r}_l}
    \hat{c}^\dagger_{l,\boldsymbol{r}_l,\uparrow}
    \hat{c}_{l,\boldsymbol{r}_l,\uparrow}
    \hat{c}^\dagger_{l,\boldsymbol{r}_l,\downarrow}
    \hat{c}_{l,\boldsymbol{r}_l,\downarrow}
    .
\end{align}
Here, $t$ is the hopping amplitude, which we set $t=1$ throughout the paper, for simplicity. $\boldsymbol{a}_{l}$ is the lattice constant of layer $l$ and $\boldsymbol{b}_l$ is its reciprocal vector.
As an interlayer coupling, we consider the following form,
\begin{align}\label{eq : interlayer coupling numerical calculation}
    \hat{V} =& \sum_{\boldsymbol{r}_1,\boldsymbol{r}_2,\sigma} \tilde{V}(|\boldsymbol{r}_1-\boldsymbol{r}_2|) \hat{c}_{\boldsymbol{r}_1,\sigma}^\dagger \hat{c}_{\boldsymbol{r}_2,\sigma},
    \\
   \tilde{V}(r) =& \begin{cases}
       V_0 e^{-r} & r \leq R \\
       0 & r > R, 
   \end{cases}  
\end{align}
with the parameter $R$ that restricts the range of the interlayer coupling.
We note that we neglected the effect of the interlayer distance and replaced $|\cdots|$ with the one-dimensional Euclidean distance in the above equation.
Since we discuss the effect of the number of folding on the SC gap function in Fig.~\ref{fig : Convergence against the folding incommensurate}, we explicitly write the BZ folding as $\boldsymbol{B}_1(F) = \{ b_1 f | f = -F,-F+1,\ldots,F \}$
with the cutoff parameter $F\in \mathbb{N}$ and $\boldsymbol{B}_2(F) = \{ fb_2 | f = -F,-F+1,\ldots,F \}$.
When we specify the cutoff parameter $F$, we can also specify the moir\'e BdG Hamiltonian $H_{BdG,\boldsymbol{B}_2(F),\boldsymbol{B}_1(F)}(\boldsymbol{k})$. The SC gap function of this BdG Hamiltonian and its elements are denoted as $\Delta(F)$ and $\Delta(F)_{l,\boldsymbol{q}}$ to emphasize $F$ dependence.

\begin{figure}[t]
    \centering
    \includegraphics[width = \linewidth]{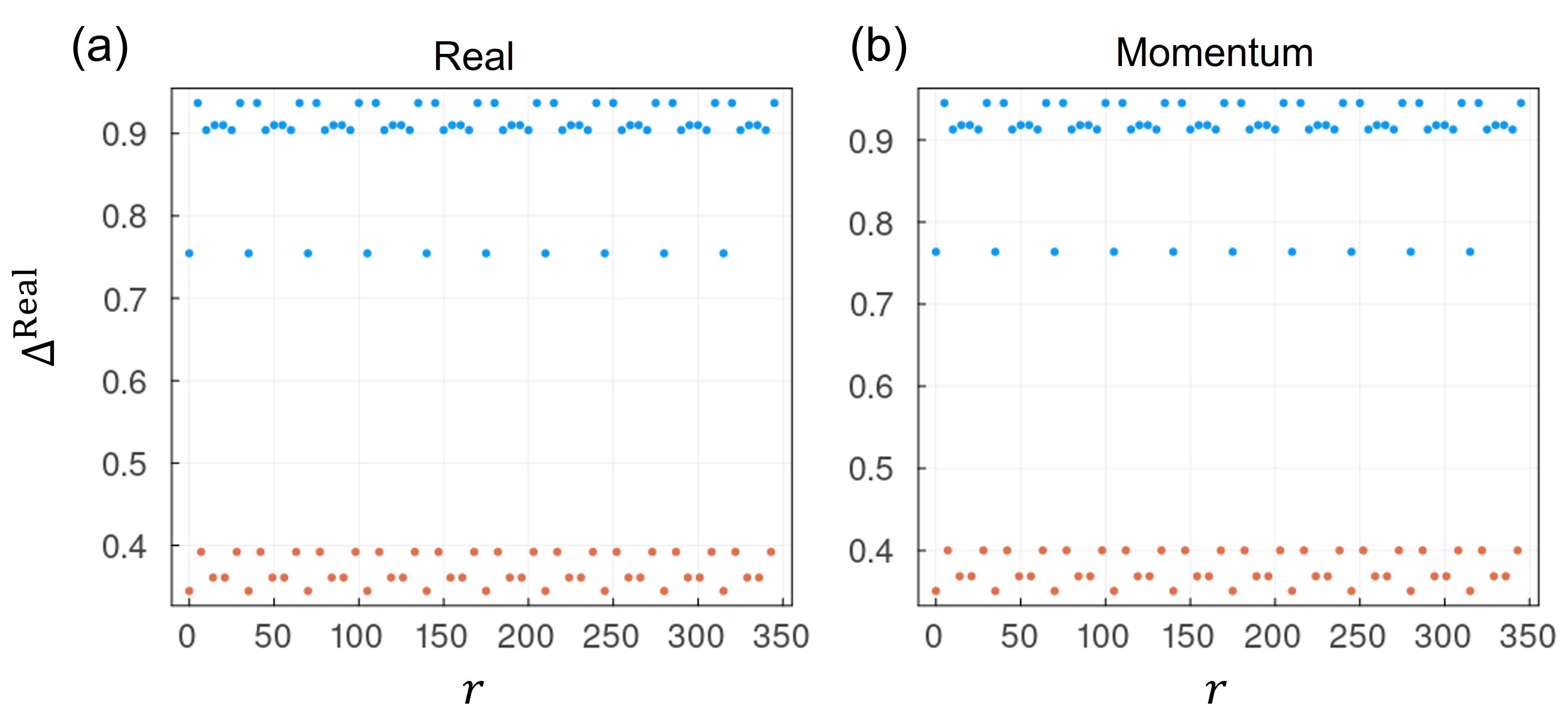}
    \caption{
    Spatial dependence of the SC gap function in the commensurate bilayer system with the lattice constants $(a_1,a_2) = (5,7)$. Light blue dots show the SC gap function on layer $1$ $[\Delta^{real}_1(\boldsymbol{r})]$, and orange dots show the SC gap function on layer $2$ $[\Delta^{real}_2(\boldsymbol{r})]$.
    (a) The SC gap function calculated from the real space approach under the periodic boundary conditions (PBCs) of the period $350$.
    (b) SC gap function calculated from the BZ folding method. The dimension of the folded Hamiltonian for this system is $12$.
    We adopted the parameters:
    $(g_1,g_2) = (3,2)$, $V_0=1$, $\beta = 1000$ and $R = 3$.
    }
    \label{fig : OP_commensurate}
\end{figure}

\subsection{Gap functions}
First, we study the SC gap functions for the commensurate case using the real space approaches and the BZ folding method, and demonstrate the validity and efficiency of our BZ folding method.
In Fig.~\ref{fig : OP_commensurate}, we show the SC gap function for the commensurate model $(a_1,a_2) = (5,7)$. The dimension of the normal state Hamiltonian is $120$ in the real space approach, whereas the dimension is $12$ in the BZ folding method. While the dimension of the momentum Hamiltonian is more than $10$ times smaller than that in the real space, we have the same result.

\begin{figure}[t]
    \centering
    \includegraphics[width = \linewidth]{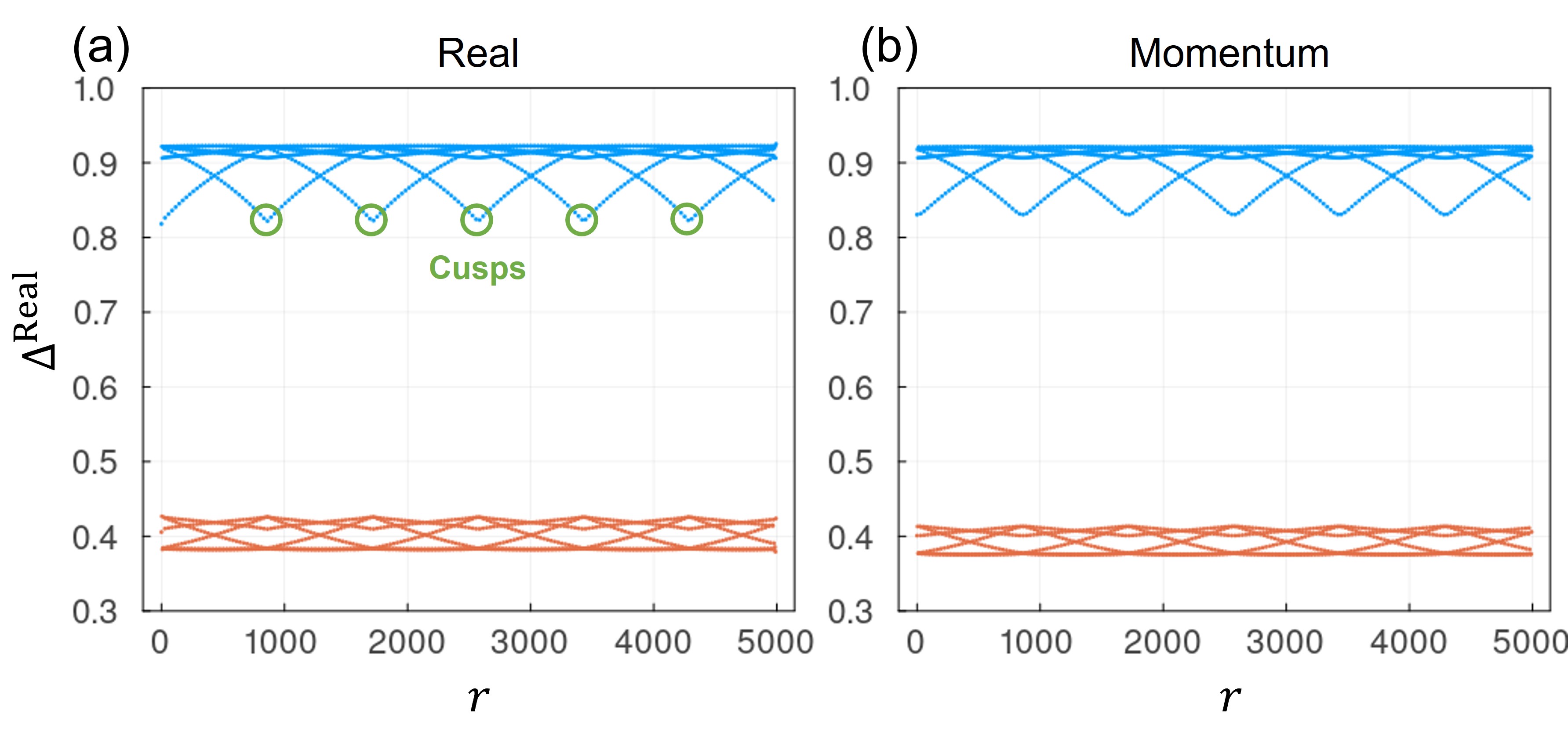}
    \caption{
    Spatial dependence of the SC gap function in the incommensurate bilayer system with the lattice constant $(a_1,a_2) = (\sqrt{13},\sqrt{17})$.
    Light blue dots show the SC gap function on layer $1$ $[\Delta^{real}_1(\boldsymbol{r})]$, and orange dots show the SC gap function on layer $2$ $[\Delta^{real}_2(\boldsymbol{r})]$.
    (a) The SC gap function calculated from the real space approach under the PBCs of period $5000$.
    (b) The SC gap function calculated from the BZ folding method. Here we set $F=32$. We used the $k$ integration area of $\boldsymbol{k} \in[-2\pi, 2\pi]$.
    The BZ folding method reproduces the overall spatial dependence of the SC gap function well. 
    One noticeable difference between (a) and (b) is that, the cusps for layer $1$ highlighted green circles in (a) are rounded in (b).
    We adopted the parameters $(g_1,g_2) = (3,2)$, $V_0=1$, $\beta = 100$ and $R=3$.
    }
    \label{fig : OP_incommensurate_real}
\end{figure}

\begin{figure*}[t]
    \centering
    \includegraphics[width = \linewidth]{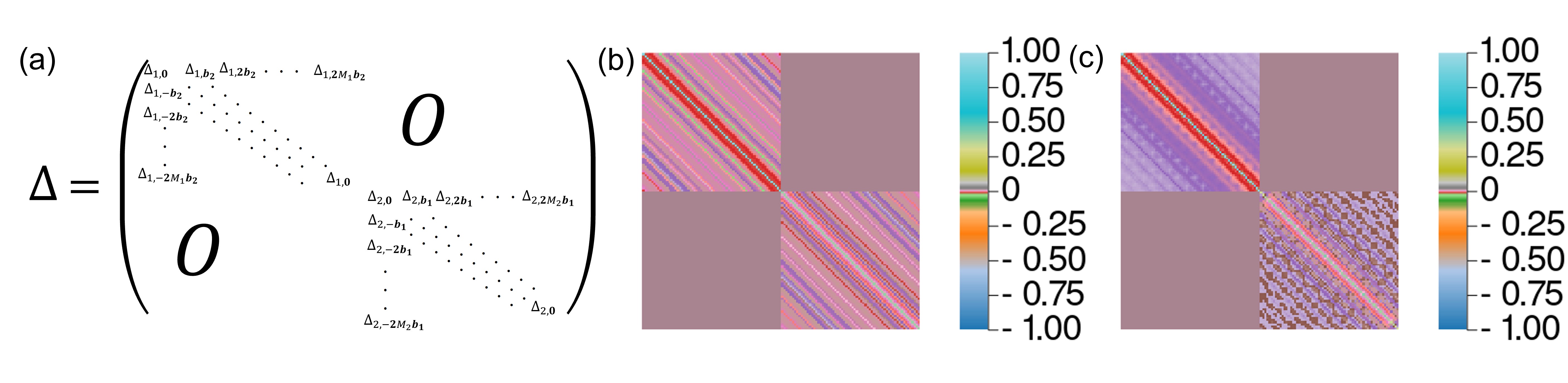}
    \caption{
    Detailed view of the SC gap function in the matrix form for the same parameters as in Fig.\ref{fig : OP_incommensurate_real}. 
    (a) The expression of the SC gap function for the one-dimensional bilayer system in the matrix form [Eq. \eqref{eq : order parameter in matrix form}].
    The top left block is the SC gap function for layer $1$ and the bottom right block is the SC gap function for layer $2$.
    In each block, $n$th diagonal elements have the same Cooper pair momentum.
    As $|n|$ grows, the norm of the Cooper pair momentum $|\boldsymbol{q}|$ increases. 
    (b) Color plot of the SC gap function $\Delta$ in the matrix form calculated with the real space approach. 
    We used the system size of 5000 and adopted the periodic boundary condition.
    After calculating the SC gap function from the real space, we apply the Fourier transformation to match the expression in panel (a).
    Due to this procedure with the Fourier transformation, diagonal elements have exactly the same values.
    (c) Color plot of the SC gap function $\Delta$ in the matrix form calculated with the BZ folding method with $F=32$.
    Matrix elements of small Cooper pair momentum have values close to those in (b), while the deviation from (b) becomes more significant for larger Cooper pair momenta (larger $|n|$).
    Also, the elements in the same diagonal part (the same $n$) take different values due to the fact that each element is calculated from electron pairs with different folding vectors.
    We adopted parameters: $(a_1,a_2) = (\sqrt{13},\sqrt{17})$, $(g_1,g_2) = (3,2)$, $V_0=1$, $\beta = 100$ and $R = 3$.
    }
    \label{fig : OP_incommensurate_momentum}
\end{figure*}

Next, we study the incommensurate cases, and show that the SC gap function is well described by our momentum approach. 
In Fig.~\ref{fig : OP_incommensurate_real} and ~\ref{fig : OP_incommensurate_momentum}, we set lattice constants to be $(a_1,a_2) = (\sqrt{13},\sqrt{17})$, and the other parameters to be $V_0 = 1$, $\beta=100$ and $R=3$.
In the real space calculation, we take the periodic boundary conditions (PBCs) with the system size $5000$.
In the momentum space calculation, we set $F=32$ and performed $\boldsymbol{k}$ integration over the area from $\boldsymbol{k} = -2\pi$ to $\boldsymbol{k} = 2\pi$.
We show the result for the spatial fluctuation of the SC gap function in Fig.~\ref{fig : OP_incommensurate_real}. Figure~\ref{fig : OP_incommensurate_real}(a) is the SC gap function directly obtained by the real space calculation, whereas
Fig.~\ref{fig : OP_incommensurate_real}(b) is the SC gap function calculated from the BZ folding method. 
In this calculation, the dimension of the real space Hamiltonian is $2598$ while the dimension of folded Hamiltonian is $130$. Despite the significantly smaller size of the folded Hamiltonian, the BZ folding method can reproduce the overall spatial dependence of the SC gap function very well.
Unlike the commensurate case, we can find some difference between real and momentum space calculation; in Fig.~\ref{fig : OP_incommensurate_real}(a), there are cusps on the SC order parameter on layer $1$, while they are rounded in Fig.~\ref{fig : OP_incommensurate_real}(b).  
Since the spatial fluctuation of the SC gap function and the finite momentum Cooper pairs are related under the Fourier transformation, we can regard it as the Gibbs phenomenon in the Fourier transformation.
This is why, we need more foldings to reproduce these cusps.
All that said, to discuss the phase transition, long range order plays an important role and this method can efficiently capture this long range order.

Next, let us look at the details of the SC gap functions obtained by the two approaches in the matrix form [Eq.~(\ref{eq : order parameter in matrix form})] in Fig.~\ref{fig : OP_incommensurate_momentum}. This figure essentially describes the gap functions in the momentum space representation with the BZ folding.
Figure~\ref{fig : OP_incommensurate_momentum}(a) reproduces the definition of the matrix form of the gap function [Eq.~(\ref{eq : order parameter in matrix form})] for the ease of interpreting the data shown in Fig.~\ref{fig : OP_incommensurate_momentum}(b) and (c).
The top left block corresponds to the SC gap function on layer $1$, and the bottom right block corresponds to the SC gap function on layer $2$.
If we call the elements $a_{i,j}$ satisfying $i-j = n$ as ``the $n$th diagonal part'',
in each block, $0$th diagonal part of each block corresponds to zero momentum Cooper pairs on each layer
The $n_{\geq 1}$th diagonal part corresponds to the finite momentum Cooper pairs,
where the $n$th diagonal part represents the Cooper pairs of the same momentum $\boldsymbol{q}$ and the norm of the Cooper pair momentum $|\boldsymbol{q}|$ increases with $|n|$.

Fig.~\ref{fig : OP_incommensurate_momentum}(b) is the SC gap function obtained by the real space calculation and Fig.~\ref{fig : OP_incommensurate_momentum}(c) is the SC gap function obtained by the BZ folding method.
In this plot, we scale the color code to vary well around $\Delta = 0$.
As we can see, the off-diagonal elements of $\Delta$ take finite value. These elements are the finite momentum Cooper pairs in the extended momentum space and describe spatial fluctuation of the SC gap function.
In small $|n|$, the $n$th diagonal region of the SC gap function in Fig.~\ref{fig : OP_incommensurate_momentum}(b) and (c) agree well. 
Since small $|n|$ corresponds to small $|\boldsymbol{q}|$,
this also means the overall behavior of real and momentum space calculations agree well.
In contrast, SC gap functions of large $|n|$ show different behaviors for (b) and (c). Since large $|n|$ corresponds to large $|\boldsymbol{q}|$, this difference means the microscopic behavior of the SC gap function is not fully reproduced by the BZ folding method.

As explained in Fig.~\ref{fig : OP_incommensurate_momentum}(a), ideally, each elements in $n$th diagonal part of $\Delta$ should be equal, but not in Fig.~\ref{fig : OP_incommensurate_momentum}(c).
This is due to the fact that each $\Delta^{l}_{\boldsymbol{b}}$ is calculated from different pairs of electrons of folded bands.
For example, $\Delta_{1,\boldsymbol{b}_2}$ can be defined as $\Delta_{1,\boldsymbol{b}_2} = -\frac{g_1}{N_1}
    \sum_{\boldsymbol{k}^\prime}
    \langle 
    \hat{c}_{1,\boldsymbol{b}_2-\boldsymbol{k}^\prime} \hat{c}_{1,\boldsymbol{k}^\prime} \rangle$,
    while it can also be defined as
    $\Delta_{1,\boldsymbol{b}_2} = -\frac{g_1}{N_1}
    \sum_{\boldsymbol{k}^\prime}
    \langle 
    \hat{c}_{1,2\boldsymbol{b}_2-\boldsymbol{k}^\prime} \hat{c}_{1,-\boldsymbol{b}_2+\boldsymbol{k}^\prime} \rangle$. 
Ideally, these two expressions should be equivalent, but not in practical calculations. Since we pick up only a finite number of BZ foldings in the numerical calculation, this makes the two expressions inequivalent.
In practice, we may average $n$th diagonal components to obtain the gap function.

\subsection{Convergence behavior}
\begin{figure}[t]
    \centering
    \includegraphics[width = \linewidth]{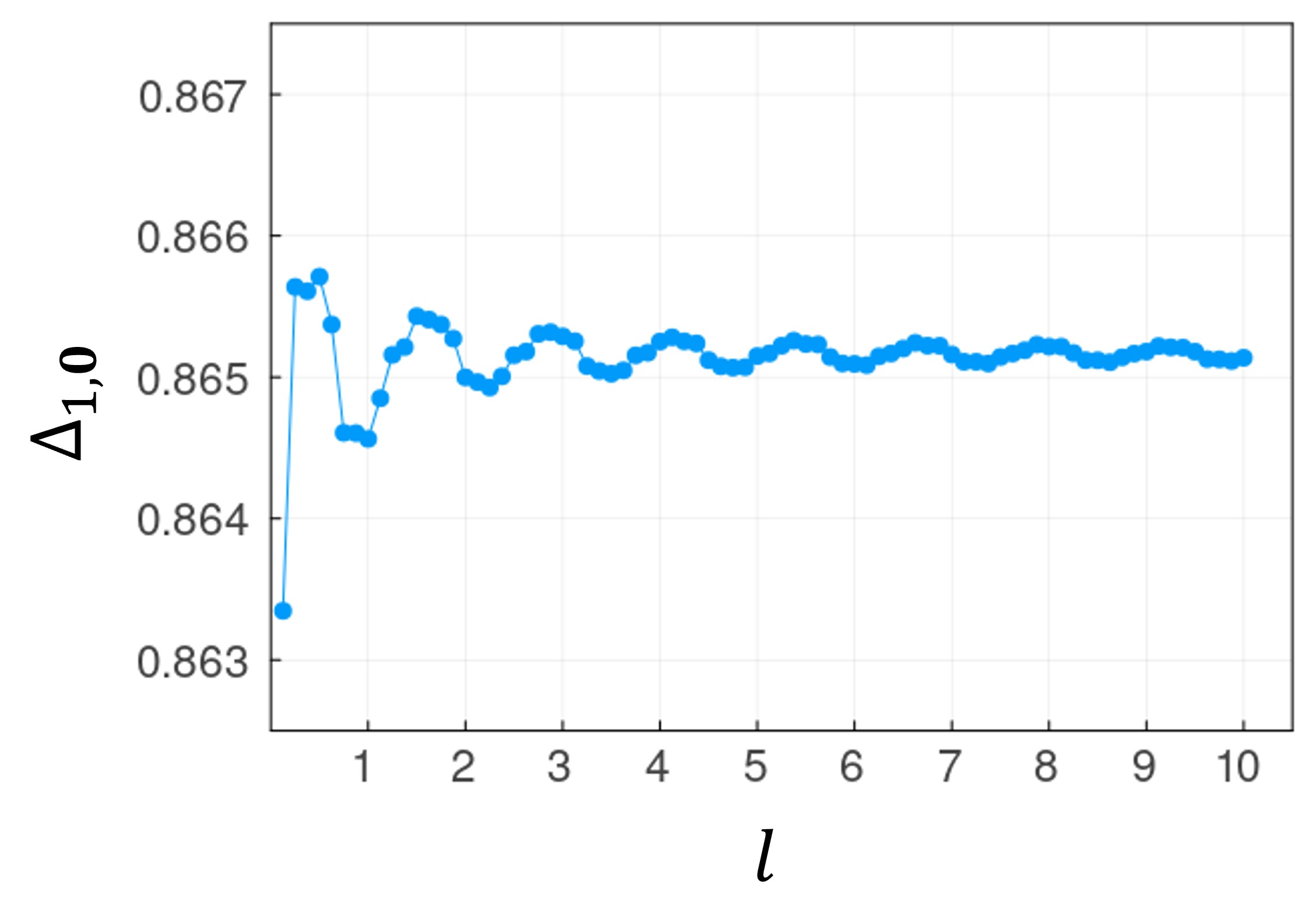}
    \caption{
    Convergence behavior of the gap function with respect to the size of the integration area $k \in [-2\pi l, 2\pi l]$.
    As a representative  matrix element, we show the data for $\Delta_{1,\boldsymbol{0}}$.
    In the self-consistent calculation, the integration is conducted over the area of $k=-2\pi l$ to $k=2\pi l$ with a $k$ mesh of $1/256$.
    We adopted the parameters to be $(a_1,a_2) = (\sqrt{13}, \sqrt{17})$, $V_0 = 1$, $\beta = 100$, $R = 3$ and $F=32$.
    }
    \label{fig : Error_vs_range}
\end{figure}

In this section, we study the convergence behaviors of the SC gap function with respect to the integration area and the number of BZ foldings.
In Fig.~\ref{fig : Error_vs_range}, we show the error against the integration area.
Here, we consider $(a_1,a_2) = (\sqrt{13}, \sqrt{17})$, $V_0 = 1$, $\beta = 100$, $R = 3$ and $F=32$. 
In this system, we parameterize the integration area to be $[-2\pi l,2\pi l]$.
As we perform the $\boldsymbol{k}$ integration with a larger $l$, the SC gap function converges.
When we increase the integration area by changing $l$ from $0.125$ to $10$,
sampled $k$-points cover two-dimensional torus densely as explained in Appendix \ref{sec : Integration over a large area}. Correspondingly, $\Delta_{1,\boldsymbol{0}}$ converges to $\Delta_{1,\boldsymbol{0}} \simeq 0.865$ within an error of $10^{-3}$.
We note that compared to the BZs of each layer $(2\pi/\sqrt{13}\approx 1.74, 2\pi/\sqrt{17} \approx 1.52)$, the integration area of $k$ for $l=10$ is approximately $40$ times larger than them.

\begin{figure}[t]
    \centering
    \includegraphics[width = \linewidth]{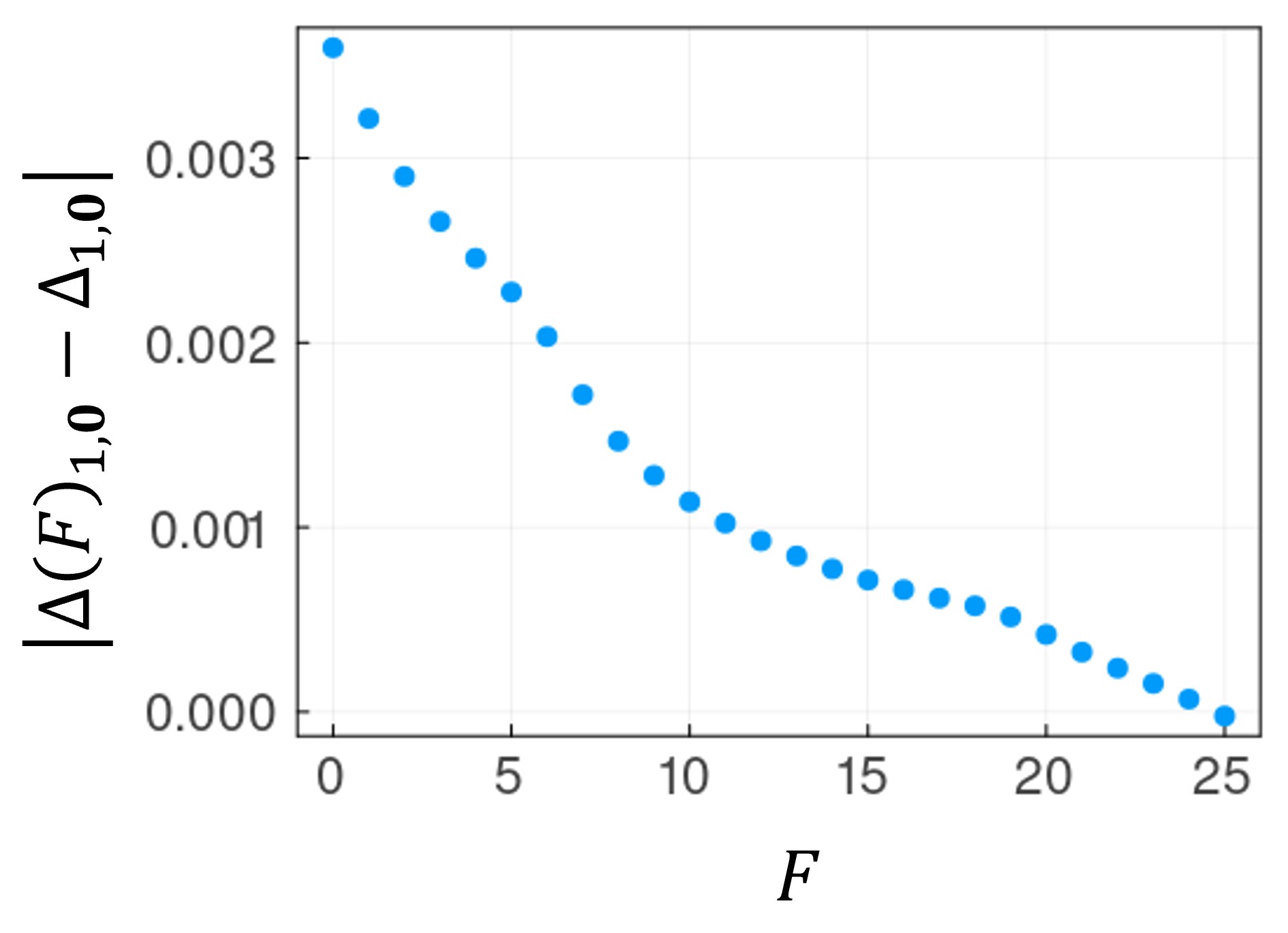}
    \caption{
    Convergence of the SC gap function.
    The lattice constants are $(a_1,a_2) = (51,53)$, and we set $V_0 = 1$, $\beta = 100$ and $R=3$. We select the folding to be $\boldsymbol{B}_1 = \{ f\boldsymbol{b}_1 | f = -F,\ldots,F \}$ and $\boldsymbol{B}_2 = \{ f\boldsymbol{b}_2 | f = -F,\ldots,F \}$.
    The vertical axis is the difference between the $\Delta_{1,\boldsymbol{0}}$ calculated by this approach and $\Delta_{1,\boldsymbol{0}}$ calculated in the conventional approach.
    }
    \label{fig : Convergence against the folding commensurate}
\end{figure}

\begin{figure*}[t]
    \centering
    \includegraphics[width = \linewidth]{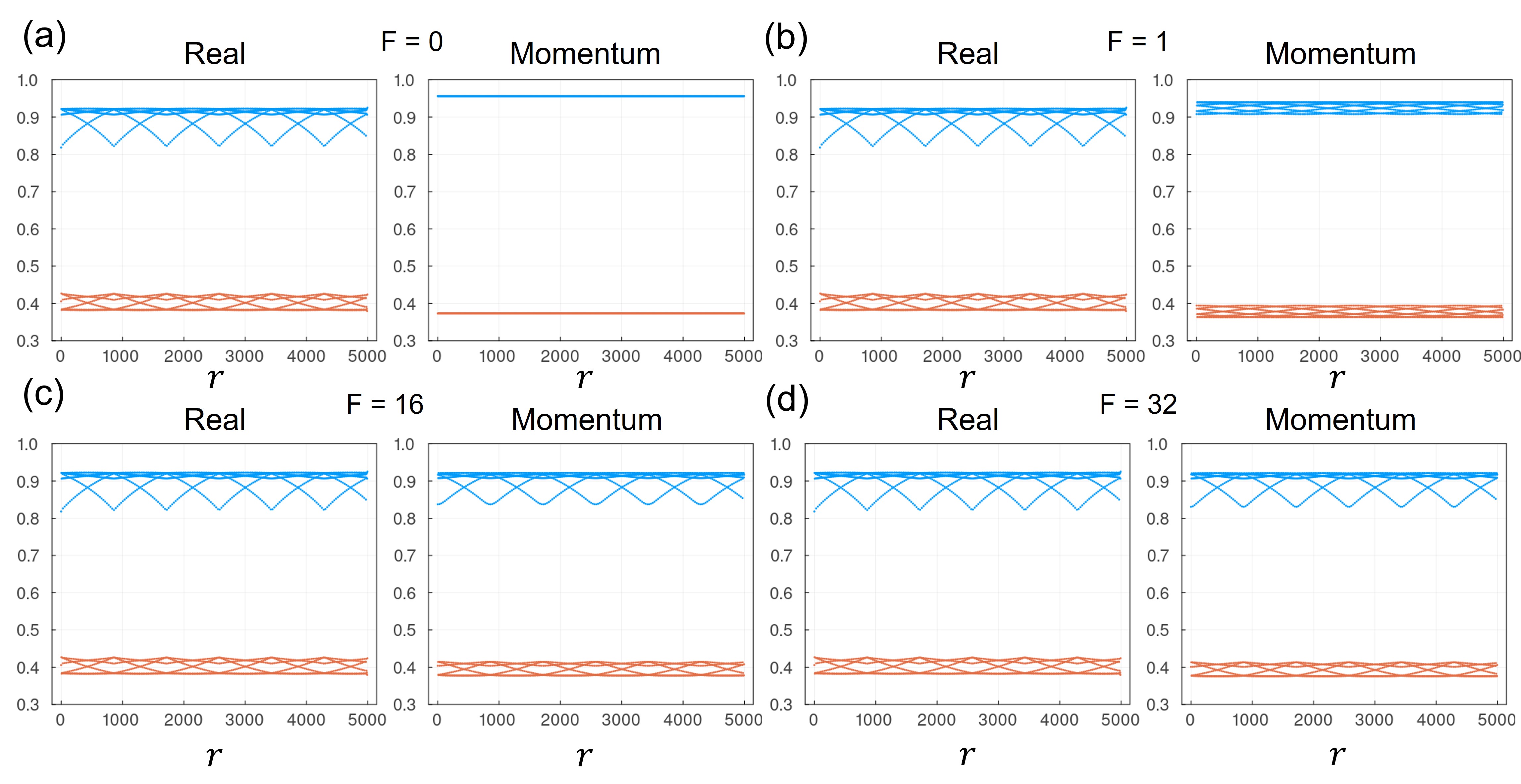}
    \caption{
    The SC gap function $\Delta(F)$ for different BZ folding $F$ compared with the real space calculation.
    (a) $F=0$.
    (b) $F=1$.
    (c) $F=16$.
    (d) $F=32$.
    The SC gap function is plotted in the real space representation, where the horizontal axis is the position in the real space from 0 to 5000. 
    For the real space calculation, we used the system with the size $5000$. 
    Without BZ folding, $F = 0$, the SC gap function is a constant function.
    For $F = 1$, the SC gap function shows a spatial dependence.
    While the overall behavior of the SC gap function is not reproduced when $F=1$, it is reproduced well in $F=16$ and $F = 32$.
    Moreover, as we increase the folding from $16$ to $32$, the cusps for blue curves become sharper. This means microscopic structure can also be reproduced once we include enough BZ foldings.
    }
    \label{fig : Convergence against the folding incommensurate}
\end{figure*}

Next, we study the convergence behavior with respect to the folding number $F$ in commensurate and incommensurate systems (Fig.~\ref{fig : Convergence against the folding commensurate} and Fig.~\ref{fig : Convergence against the folding incommensurate}).
Figure~\ref{fig : Convergence against the folding commensurate} shows the convergence of the SC gap function $\Delta_{1,\boldsymbol{0}}(F)$ in the commensurate system with the lattice constants $(a_1,a_2) = (51,53)$, obtained from the BZ folding method with folding $F$.
Since we can use the conventional real space approach for commensurate systems, we show the deviation $|\Delta_{1,\boldsymbol{0}}(F)-\Delta_{1,\boldsymbol{0}}|$ from the exact value $\Delta_{1,\boldsymbol{0}}$ obtained by the conventional approach,
where one can see the SC gap function $\Delta_{1,\boldsymbol{0}}(F)$ is quickly converging to the exact value $\Delta_{1,\boldsymbol{0}}$.

Finally, we show the SC gap function $\Delta(F)$ in the incommensurate system for different values of $F$ in Fig.~\ref{fig : Convergence against the folding incommensurate},  demonstrating the convergence behavior of the SC gap function with $F$. 
Since we cannot define superlattice structure in the incommensurate case, we used a sufficiently large system with the size $5000$ in the real space calculation, for which the dimension of the Hamiltonian is $2598$.
In the BZ folding method, we set the integration area to be $l=1$
and varied the BZ folding from $F=0$ to $F=32$.
Without BZ folding, $F = 0$, the SC gap function is a constant function.
For $F = 1$, the SC gap function already shows a spatial dependence.
The overall behavior of the SC gap function is reproduced well in $F=16$ and $F = 32$.
Moreover, as we increase the folding from $16$ to $32$, the cusp (the structure highlighted with circles in Fig.~\ref{fig : OP_incommensurate_real}) becomes sharper. This means the microscopic structure can also be reproduced once we include enough BZ foldings.

\section{Shift current}\label{sec : Shift current}

\begin{figure}
    \centering
    \includegraphics[width = \linewidth]{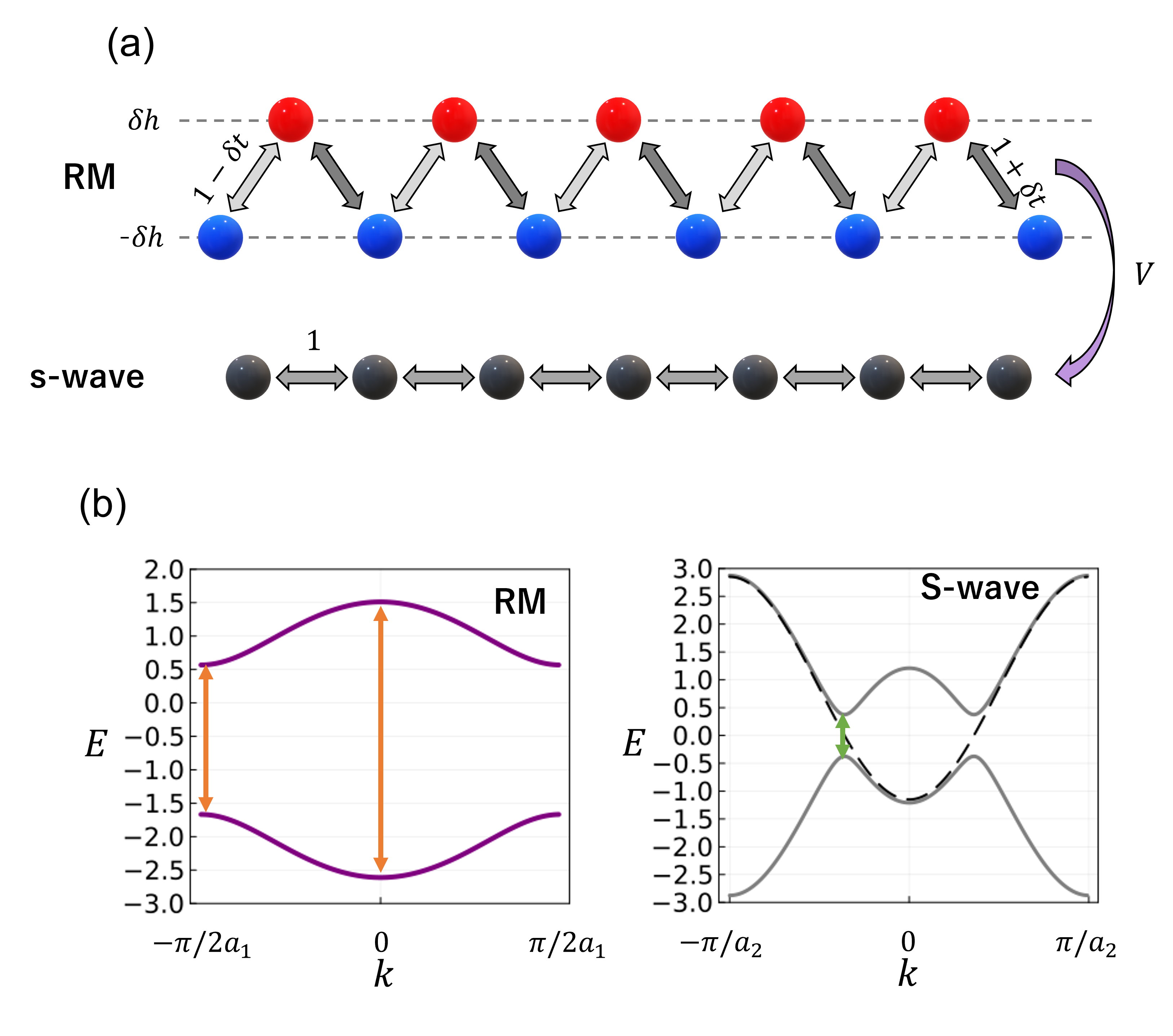}
    \caption{
    A bilayer system of the Rice-Mele (RM) model and an s-wave superconductor.  
    (a). Schematic illustration of the bilayer system consisting of the inversion broken RM model and the s-wave superconductor. 
    Black sites consists of the s-wave SC and red and blue consist of RM model.
    In RM model, we have two sublattices (blue and red), and on-site potential for red (blue) is $\delta h$ ($-\delta h$). 
    (b) Band structures of RM model (left) and the s-wave SC (right).
    Minimum and maximum of the energy separation of the two bands are indicated by orange arrows. 
    Dashed curves in the right panel shows the energy band in the normal state, and the solid curves show the energy band in the SC state.
    }
    \label{fig : RM + Swave}
\end{figure}
As an application of the present method, we demonstrate the numerical calculation for the optical responses 
in the bilayer system.
Especially, we focus on the optical absorption and the shift current here.
Shift current is a DC current response induced by the second order optical response in inversion broken materials \cite{Baltz,Sipe,Young-Rappe,Morimoto-Nagaosa16,Cook2017,Morimoto-JPSJ23}.

To study the shift current response in a quasiperiodic SCs, we consider a heterostructure of the Rice-Mele (RM) model 
\begin{align}
    \hat{H}_\mathrm{RM} = \sum_{n \in \mathbb{Z} } \sum_{\sigma} 
    &\Big\{\left[
    (1+(-1)^n\delta t)\hat{c}^{\dagger}_{a_1(n+1),\sigma} \hat{c}_{a_1n,\sigma} + h.c. \right]
    \nonumber \\
    &+\left(-\mu_1+(-1)^n \delta h\right)\hat{c}^{\dagger}_{a_1n,\sigma} \hat{c}_{a_1n,\sigma}
    \Big\},
\end{align}
and a 1D chain with the single cosine band with an s-wave superconducting order,
\begin{align}
    \hat{H}_\text{s-wave} &= \MY{-}\sum_{n \in \mathbb{Z} } \sum_{\sigma} 
    (\hat{c}^{\dagger}_{a_2(n+1),\sigma} \hat{c}_{a_2n,\sigma} + h.c. - \mu_2 \hat{c}^{\dagger}_{a_2n,\sigma} \hat{c}_{a_2n,\sigma})
    \nonumber\\
    &- g \sum_{n \in \mathbb{Z} } \hat{c}^\dagger_{a_2n,\uparrow}\hat{c}_{a_2n,\uparrow}
    \hat{c}^\dagger_{a_2n,\downarrow}\hat{c}_{a_2n,\downarrow}.
\end{align}
These models are coupled with the interlayer coupling defined in Eq.~\eqref{eq : interlayer coupling numerical calculation}.  When both $\delta t$ and $\delta h$ are nonzero, the RM model breaks the inversion symmetry and introduces an inversion symmetry breaking for the s-wave SC through the interlayer coupling.
The sublattice degrees of freedom in the RM model is incorporated by treating them as additional layer degrees of freedom as announced in the beginning of Sec.~\ref{sec : Formalism}.
Specifically, we regard the sites of odd $n$ and even $n$ as different layers as illustrated in Fig.~\ref{fig : RM + Swave}~(a). In this picture, there is no intralayer hopping and only the interlayer hopping,
\begin{align}
    \hat{V}_\mathrm{RM} = \sum_{n \in 2\mathbb{Z}}\left[
    (1+\delta t)\hat{c}^{\dagger}_{n+1} \hat{c}_{n}
    +
    (1-\delta t)\hat{c}^{\dagger}_{n-1} \hat{c}_{n}
    + h.c.
    \right],
\end{align}
along with the on-site potential $\pm \delta h$ for two layers.
We note that, to obtain physical quantities, we use the integration area for the $k$-space as $\boldsymbol{k} \in [-\pi, \pi]$ since we cannot define the BZ in the incommensurate system, 

Figure~\ref{fig : Shift current} shows the numerical result for the optical response obtained using the formulas in Appendix \ref{sec : Optical response}.
Figures~\ref{fig : Shift current}(a) and (b) show nonlinear conductivity $\kappa^{xxx}$ of the shift current for commensurate and incommensurate cases, and (c) and (d) show the corresponding optical absorption $\alpha$.
We use the lattice constant $(a_1,a_2) = (3,5)$ and $(a_1,a_2) = (3,\sqrt{24})$ for the commensurate and incommensurate systems, respectively.
First, the gray dashed curve in each panel show the result when the interaction and the interlayer coupling are turned off. In this case, the cosine band does not contribute to the optical response as far as we focus on the interband transition, and the optical responses takes place in the region of interband transition of the RM model which is highlighted in orange color in each panel.
Second, light-blue curves show the result with finite interlayer coupling and zero interaction term $(V \neq 0, g = 0)$.
We find that the finite interlayer coupling gives rise to optical transitions within the 1D chain with the cosine band and between the two layers, leading to nonzero signals below the band gap of the RM model (below the orange region).
Finally, let us look at the magenta curves with finite interlayer coupling and interaction term $g = 2$.
With the contact interaction, we have a strong shift current response at the superconducting gap below the band gap of the RM model in addition to the interband transition of the RM model above the band gap.
In particular, we have a strong peak around $\omega = 0.75$, which corresponds to $2\Delta_{2,0}$ as indicated by the green line in the insets.
This shift current response at the SC gap arises from the photoexcitation of a Bogoliubov quasiparticle in the s-wave superconductor which accompanies nonzero electric polarization due to the coupling to the inversion broken RM model.

\begin{figure*}
    \centering
    \includegraphics[width = 0.8\linewidth]{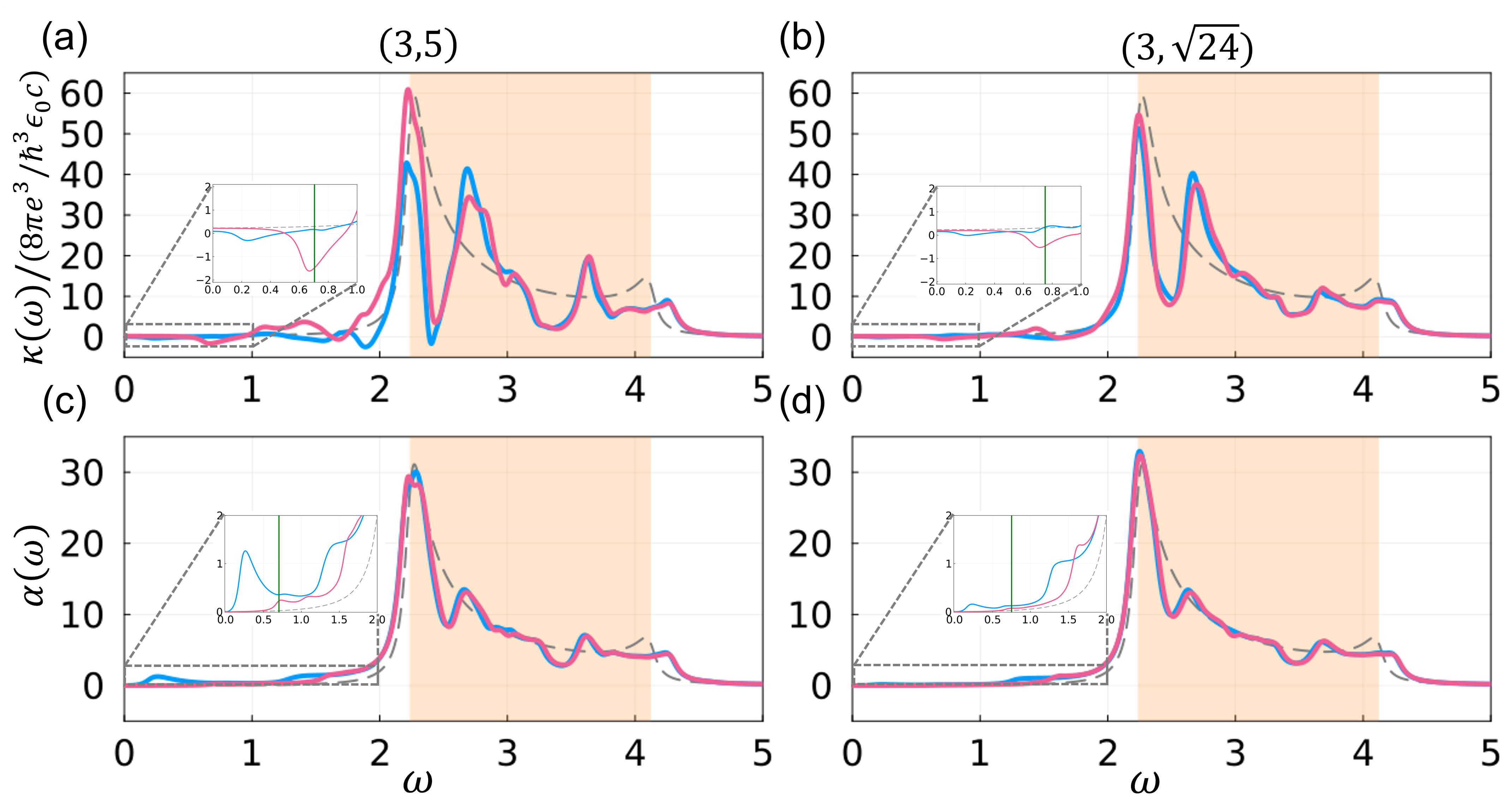}
    \caption{
    Optical responses in the bilayer system of the Rice-Mele (RM) model and an s-wave superconductor with and without the contact interaction.
    (a,b). Nonlinear conductivity $\kappa^{xxx}$ plotted against the frequency of input light $\omega$ for (a) commensurate and (b) incommensurate systems.
    (c,d). Optical absorption plotted against the frequency of input light $\omega$ for (c) commensurate and (d) incommensurate systems.
    Gray dashed curves show the case where the interaction and the interlayer coupling are turned off. In this case, the cosine band does not contribute to the optical response, and the optical responses takes place in the region of interband transition of the RM model which is highlighted in orange color in each panel.
    Light-blue curves represent the shift current without the contact interaction.
    Due to the interlayer coupling, the interlayer transition takes place between the single cosine band and two bands from the RM chain, in addition to an intralayer optical transition.
    Magenta curves represent the shift current with the contact interaction.
    With the contact interaction, we have a photoexcitation of the Cooper pairs.
    In shift current calculation, we have large peaks which correspond to the SC gap indicated by the green lines corresponding to $\hbar\omega = 2\Delta_{2,0}$.
    In the optical absorption, we can observe the kink corresponding to the SC gap.
    In this calculation, we fold the band from the RM model by 9 times and the band from s-wave SC by 11 times.
    We use  the lattice constant $(a_1,a_2) = (3,5)$ for the commensurate systems and  $(a_1,a_2) = (3,\sqrt{24})$ for the incommensurate systems.
    The k-integration is taken over $\boldsymbol{k} \in [-\pi, \pi]$.
    We used the parameters, $(\mu_1,\mu_2) = (0.55,0.85),g = 2, R = 3, \beta = 2^{20}, \delta t = 0.5, \delta h = 0.5$.
    }
    \label{fig : Shift current}
\end{figure*}

\section{Discussion}\label{sec : Discussions}
In this paper, we extended the BZ folding method to the SC phase and relate the spatial fluctuation of the SC gap function to the finite momentum Cooper pairs in the extended space.
To this end, the conventional folded Hamiltonian in the normal state is extended to the BdG Hamiltonian form by using the particle-hole symmetry.
Specifically, in the Nambu space representation for the extended momentum space, the particle-hole symmetry indicates that the hole part has the BZ folding vectors of opposite signs with respect to those for the electron part, in addition to the sign flip of the wave number $\boldsymbol{k}$, as in Eq.~(\ref{eq : Nambu spinor extended momentum sp}).
In this approach, the main difference between the quasiperiodic systems and the periodic systems comes from the BZ foldings.
Namely, Cooper pairs introduced by the BZ folding may possess a finite center of mass momentum, and corresponds to the spatial fluctuation of the SC order parameter.

In practical calculations, we apply two approximations that result in some difference between the real space approach and the BZ folding method.
One is for the BZ folding.
Since the spatial fluctuation of the SC gap function and the finite momentum Cooper pair are related under the Fourier transformation, we need many foldings to describe sharp structures.
All that said, to discuss the phase transition, long range order plays an important role and this method can efficiently capture this long range order.
The second approximation is for the integration area.  
The integration of the superconducting pairing amplitude over the infinitely large area in the incommensurate system is approximated with that over a large area in the practical calculations. The SC gap function calculated under this approximation converges to the result from the real space approach in the thermodynamic limit as the integration area increases.
Since these two approximations are controllable with the choice of BZ folding and integration area, we can choose them according to the required precision, for example, by the corresponding experimental accuracy.
Thus, the present BZ folding method provides an efficient approach to compute superconducting orders in quasiperiodic systems.

Finally, we mention the possible future directions to extend the present method.
In this paper, we discussed only the contact interaction.
This is a minimal model which takes into account both quasiperiodicity and superconductivity.
This approximation also plays a role in reducing the computational cost.
In general, the interaction strength has a dependence on the relative momentum of holes and electrons $(\boldsymbol{k})$,
which leads to more computational cost. 
In addition, for general interaction, the center of mass momenta of Cooper pairs do not conserve and we need to consider momentum transfer.
One possible application of the present method is the large angle twisted bilayer (either homolayer or heterolayer) systems that do not have interlayer Cooper pairing.
In particular, since not all layers have to be superconducting in the present method,
the heterolayer systems made of a superconductor and a normal (semi)conductor can be studied with the BZ folding method, 
as demonstrated in Sec.~\ref{sec : Shift current}. 
While we employed RM model to introduce the inversion symmetry breaking to the SC, we may also consider other systems as the stacking layer, such as a ferromagnet and a topological insulator, which is expected to provide novel functionalities in superconducting MLTF.

\acknowledgements
We thank Takami Tohyama, Kaori Tanaka and Masahiro Hori for fruitful discussions.
This work was supported by 
JSPS KAKENHI Grant 23KJ0551 (M.Y.),  20K14407 (S.K.), 23K25816, 23K17665, 24H02231 (T.M.), and
JST CREST (Grant No. JPMJCR19T3) (T.M., S.K.).

\appendix

\section{Functions defined on the MLTFs}\label{sec : Functions defined on the MLTFs}
In MLTFs, physical observables which follows spatial structure of MLTFs can be related to the higher dimensional periodic function \cite{rosa2021topological,PhysRevB.107.064201}. Here, we show how to construct the corresponding function.
We consider a $D$-dimensional $L$ layer system without spontaneous spatial symmetry breaking. For simplicity, we consider a scalar valued observable $\mathcal{O}$ defined for the real space coordinate $\boldsymbol{r} \in \mathbb{R}^D$ as a function $\mathcal{O} : \mathbb{R}^D \mapsto \mathbb{R}$.

\subsection{Coordinates}
Here, we define two coordinate systems. 
One is the redundant coordinate system $\mathbb{R}^{DL}$.
Even though we assume that $\mathbb{R}^D$ coordinate system (we name multi-layer (ML)-coordinate system) is defined for the MLTF itself, we may also define $\mathbb{R}^D$ coordinate systems on each layer, which we name single-layer (SL)-coordinate system. 
When we set the origin of SL-coordinate system on layer $l$ to be $\boldsymbol{\gamma}_{ol}$ on ML-coordinate system, $\boldsymbol{\gamma}_l$ on SL-coordinate system corresponds to ML-coordinate $\boldsymbol{r}$ as
\begin{align}
    \boldsymbol{r} = \boldsymbol{\gamma}_{l} + \boldsymbol{\gamma}_{ol}.
\end{align}
$\boldsymbol{\gamma}_o = (\boldsymbol{\gamma}_{o1},\ldots,\boldsymbol{\gamma}_{oL})$ describes the sliding of layers, because if we change $\boldsymbol{\gamma}_{ol}$ by $\Delta \boldsymbol{\gamma}_{ol}$, all atoms on the layer are also be shifted by $\Delta \boldsymbol{\gamma}_{ol}$. Therefore, we call $\boldsymbol{\gamma}_o$ sliding and $\boldsymbol{\gamma}_l$ SL-coordinate.
In addition we consider another coordinate system called the unit-cell (UC)-coordinate.
Since we have unit cell on each layer, we can define the position inside the unit cell $\xi_{l,\boldsymbol{\gamma}_{ol}} : \mathbb{R}^D \mapsto [0,1)^D$ as $\xi_{l,\boldsymbol{\gamma}_{ol}} (\boldsymbol{r}) = (
\fraction{\boldsymbol{\gamma}_l\cdot \boldsymbol{b}_{l,1}}, \ldots, \fraction{\boldsymbol{\gamma}_l\cdot \boldsymbol{b}_{l,
D}})$, where $\fraction{x}$ is the fractional part of a real number $x$.
$\boldsymbol{\xi}_{l,\gamma_{ol}}$ is the position within the unit cell for each layer and defines the UC-coordinate system which corresponds to $\boldsymbol{r}$ in the ML-coordinate system.
\subsection{Injectivity from ML-coordinate to UC-coordinate}
When the multilayer system is incommensurate, $\boldsymbol{\xi}_{\boldsymbol{\gamma}_o=(\boldsymbol{\gamma}_{o1},\ldots,\boldsymbol{\gamma}_{oL})} := (\xi_{1,\boldsymbol{\gamma}_{o1}},\ldots,\xi_{L,\boldsymbol{\gamma}_{oL}})$ is an injective map.
This statement is proven as follows. When $\boldsymbol{\xi}$ is not injective, we can find $\boldsymbol{r},\boldsymbol{r}^\prime \in \mathbb{R}^D$ which shares the same $\boldsymbol{\xi}^\star =  \boldsymbol{\xi}_{\boldsymbol{\gamma}_o}(\boldsymbol{r}) = \boldsymbol{\xi}_{\boldsymbol{\gamma}_o}(\boldsymbol{r}^\prime)$.
This means $\xi_{l,{\boldsymbol{\gamma}_l}} (\boldsymbol{r}-\boldsymbol{r}^\prime) = (
\fraction{(\boldsymbol{r}-\boldsymbol{r}^\prime)\cdot \boldsymbol{b}_{l,1}}, \ldots, \fraction{(\boldsymbol{r}-\boldsymbol{r}^\prime)\cdot \boldsymbol{b}_{l,
D}}) = \boldsymbol{0}$ holds for all $l = 1,\ldots, L$.
Thus,  for each layer, we can find $n_{l,i} \in \mathbb{Z}$, which satisfies $\boldsymbol{r}-\boldsymbol{r}^\prime = \sum_{i=1}^{D} n_{l,i} \boldsymbol{a}_{l,i}$. This contradicts the assumption that the system is incommensurate. Therefore, $\boldsymbol{\xi}_{\boldsymbol{\gamma}_o}$ is an injective map in incomensurate systems.
However, $\boldsymbol{\xi}_{\boldsymbol{\gamma}_o}$ is not a surjective function, because $\mathbb{R}^{D} \not\simeq [1,0)^{DL}$ when $L \geq 2$. 
Thus, the pullback of $\mathcal{O}$ by $\boldsymbol{\xi}_{\boldsymbol{\gamma}_o}$ is a map from $\{ \boldsymbol{\xi}_{\boldsymbol{\gamma}_o}(\boldsymbol{r}) | \boldsymbol{r} \in \mathbb{R}^{D} \} \subset [0,1)^{DL}$ to $\mathbb{R}$, and not from $[0,1)^{DL} \mapsto \mathbb{R}$.

\subsection{Bijectivity between sliding and UC-coordinate}
To extend the map $\{ \boldsymbol{\xi}_{\boldsymbol{\gamma}_o}(\boldsymbol{r}) | \boldsymbol{r} \in \mathbb{R}^{D} \} \mapsto \mathbb{R}$ to $[0,1)^{DL} \mapsto \mathbb{R}$, we slide the layer by changing $\boldsymbol{\gamma}_o$. 
Although all elements in $\boldsymbol{\xi}_{{\gamma}_o}$ changes against $\boldsymbol{r}$, one can change only one element in $\boldsymbol{\xi}_{{\gamma}_o}$ by changing $\boldsymbol{\gamma}_o$ properly.
Indeed, when we slide layer $l$ parallel to the primitive vector $\boldsymbol{a}_{l,1}$ by $\alpha \boldsymbol{a}_{l,1} \ (\alpha \in \mathbb{R})$, $\xi_{l,\boldsymbol{\gamma}_{ol}} (\boldsymbol{r}) = (\xi_{l,\boldsymbol{\gamma}_{ol},1} (\boldsymbol{r}),\ldots, \xi_{l,\boldsymbol{\gamma}_{ol},D} (\boldsymbol{r}))$ changes to 
$(
\fraction{\xi_{l,\boldsymbol{\gamma}_{ol},1} (\boldsymbol{r}) - \alpha}, \xi_{l,\boldsymbol{\gamma}_{ol},2}(\boldsymbol{r}),\ldots, \xi_{l,\boldsymbol{\gamma}_{ol},D} (\boldsymbol{r}) )$. 
In a similar manner, one can change $\boldsymbol{\xi}$ freely inside $[0,1)^{DL}$. This enable us to define a map from $[0,1)^{DL}$ to $\mathbb{R}$ as $\mathcal{O}\circ \boldsymbol{\xi}^{-1}(\boldsymbol{r})$ by fixing $\boldsymbol{r}$. 
Here, $\boldsymbol{\xi}^{-1} (\boldsymbol{r}): [0,1)^{DL} \mapsto [0,1)^{DL}$ is the inverse of the map from $\boldsymbol{\gamma}_o$ to $\boldsymbol{\xi}_{\boldsymbol{\gamma}_o}(\boldsymbol{r})$.
Moreover, the choice of $\boldsymbol{r}$ does not affect this construction since $\boldsymbol{\xi}^{-1}(\boldsymbol{r})$ remains unchanged against the shift of $\boldsymbol{r}$ 
and one can absorb the change of $\boldsymbol{r}$ by the change of $\boldsymbol{\gamma}_o$ using $\boldsymbol{\xi}_{\boldsymbol{\gamma}_o}(\boldsymbol{r} + \Delta\boldsymbol{r}) = \boldsymbol{\xi}_{\boldsymbol{\gamma}_o-\Delta\boldsymbol{r}}(\boldsymbol{r})$.
In the following, we abbreviate $\boldsymbol{\xi}^{-1}(\boldsymbol{r})$ to $\boldsymbol{\xi}^{-1}$.
Since moving $\boldsymbol{r}$ is a special way to move $\boldsymbol{\gamma}_o$, $\mathcal{O}\circ\boldsymbol{\xi}^{-1}$ is an extension of $\{ \boldsymbol{\xi}_{\boldsymbol{\gamma}_o}(\boldsymbol{r}) | \boldsymbol{r} \in \mathbb{R}^{D} \} \mapsto \mathbb{R}$ to $[0,1)^{DL} \mapsto \mathbb{R}$.
Lastly, we extend the domain of $\mathcal{O}\circ \boldsymbol{\xi}^{-1}$ to $\mathbb{R}^{DL}$.
From $\fraction{1} = \fraction{0} = 0$, $\xi_{l,i} \in [0,1) \simeq S^1$, so $[0,1)^{DL} \simeq \mathbb{T}^{DL}$. Using this isomorphism, we can define $\mathcal{O}\circ \boldsymbol{\xi}^{-1}$ as a map from $\mathbb{T}^{DL}$ to $\mathbb{R}$. 
Moreover, we can also expand the torus to $\mathbb{R}^{DL}$ by omitting $\mathrm{frac}$ in the definition of $\boldsymbol{\xi}$ as $\tilde{\boldsymbol{\xi}}_{\boldsymbol{\gamma}_o} = (
(\boldsymbol{r}-\boldsymbol{\gamma}_{ol})\cdot \boldsymbol{b}_{l,1}, \ldots, (\boldsymbol{r}-\boldsymbol{\gamma}_{ol})\cdot \boldsymbol{b}_{l,
D})$.
As a result, one obtains a $DL$-dimensional periodic function $\mathcal{O}\circ \tilde{\boldsymbol{\xi}}^{-1}$. 

\section{Fourier expansion of quasiperiodic functions \label{sec : Fourier expansion of quasiperiodic functions}}

As explained in the previous section, physical observables are related to the higher dimensional periodic functions. This indicates that physical observables are given by quasiperiodic functions in MLTFs generally.
Using this property, we can restrict the center of mass momenta of the Cooper pairs.

The quasiperiodic function is a function obtained by projecting a higher dimensional periodic function.
For simplicity, we consider a one-dimensional system in this section.
Specifically, a one-dimensional quasiperiodic function $F$ can be obtained by projecting an $L$-dimensional periodic function $\tilde{F}$ as
\begin{align}
    F(r) = \tilde{F}(r+o_1,\ldots, r+o_L).
\end{align}
Here, $o_i\in \mathbb{R}\ (i \in \{1,\ldots,L\})$ is an arbitrary parameter to specify the projection and we may set $o_i=0$ without loss of generality. 
Note that $\tilde{F}$ satisfies
\begin{align}
    \tilde{F}(x_1,\ldots, x_i + a_i,\ldots, x_L) = \tilde{F}(x_1,\ldots, x_i,\ldots, x_L).
\end{align}
Defining the Fourier expansion of $F$ and $\tilde{F}$ as
\begin{align}
    F(r) =& \int_{-\infty}^{\infty} dq f_q e^{-iqr}, \\
    \tilde{F}(x_1,\ldots,x_L) =& 
    \sum_{\eta_l \in \frac{2\pi}{a_l}\mathbb{Z}}
    \left(
    \prod_l 
    e^{-i\eta_l x_l}
    \right) 
    \tilde{f}_{\eta_1,\ldots,\eta_L},
\end{align}
we may relate $f_q$ and $\tilde{f}_{\eta_1,\ldots,\eta_L}$ as
\begin{align}
    f_q =& 
    \sum_{\eta_1 \in \frac{2\pi}{a_1}\mathbb{Z}}\cdots \sum_{\eta_L \in \frac{2\pi}{a_L}\mathbb{Z}} \delta(q-\eta_1-\cdots-\eta_L)
    \tilde{f}_{\eta_1,\ldots,\eta_L}.
\end{align}
This shows that the Fourier coefficients of quasiperiodic systems take discrete values similar to periodic systems.

In MLTFs without spontaneous spatial symmetry breaking, the superconducting order parameter $\Delta^{real}_l(\boldsymbol{r})$ is quasiperiodic.
When the system is a one-dimensional bilayer system, $\Delta^{real}_l(r)$ is a projection of two-dimensional periodic function $\Delta_{2D,l}$ to one-dimensional system by
\begin{align}
    \Delta^{real}_l(r) = \Delta_{2D,l}(r,r).
\end{align}
Therefore, $\Delta_{G}$ in Eq.~(\ref{eq : Fourier of Delta}) can take nonzero value when $G \in \mathcal{B}_1+\mathcal{B}_2$.

\section{Detail of the Brillouin folding method}\label{sec : Detail of the BZFM}
\subsection{Detailed derivation of the interlayer coupling} \label{sec : Interlayer coupling}
In this section, we provide a comprehensive derivation of Eqs.~(\ref{eq : Fourier V1}), (\ref{eq : Fourier V2}) and (\ref{eq : Interlayer coupling extended}). We begin by assuming that the system is commensurate.
Namely, one can find a set of integers $n_{l,ij}$ which satisfy
\begin{align}
    \boldsymbol{A}_i =\sum_{j=1}^{D} n_{1,ij}\boldsymbol{a}_{1,j}
    =
    \sum_{j=1}^{D} n_{2,ij}\boldsymbol{a}_{2,j},
\end{align}
with $\boldsymbol{A}_1,\ldots,\boldsymbol{A}_D$ being linearly independent. 
We impose the periodic boundary conditions in this system by identifying $\boldsymbol{r}$ with $\boldsymbol{r}+\boldsymbol{A}_i$.
Then the system volume $S$ (that of parallelotope defined by $\boldsymbol{A}_1, \ldots , \boldsymbol{A}_D$) is related to the size of the unit cell $S_l$ of layer $l$ as
\begin{align} 
    S = N_1 S_1 = N_2 S_2,
\end{align}
where $N_l$ is the number of unit cells on layer $l$.

We define the Fourier transformation of $V_{\sigma_1\sigma_2}(\boldsymbol{r})$ to be 
\begin{align}
    V_{\sigma_1\sigma_2}(\boldsymbol{r})
    =&
    \frac{1}{\sqrt{N_1N_2}}
    \sum_{\boldsymbol{q}}
    v_{\sigma_1\sigma_2,\boldsymbol{q}}e^{i\boldsymbol{q}\cdot\boldsymbol{r}},
    \label{eq : fourier V with alpha}
    \\
    v_{\sigma_1\sigma_2,\boldsymbol{q}}
    =&
    \frac{1}{\sqrt{S_1S_2}}
    \int_{S} d^Dr V_{\sigma_1\sigma_2}(\boldsymbol{r})e^{-i\boldsymbol{q}\cdot\boldsymbol{r}},
\end{align}
where we have chosen the normalization factor such that the matrix element of $\hat{V}$ is given by $v$ (see below).
Here, $\boldsymbol{q}$ takes a discrete value $\boldsymbol{q} = \sum_{i=1}^D n_{i} \boldsymbol{B}_i\ (n_i \in \mathbb{Z})$ where $\boldsymbol{B}_i$ is defined as $\boldsymbol{A}_i\cdot\boldsymbol{B}_j = 2\pi \delta_{i,j}$.
Using Eq.~\eqref{eq : fourier V with alpha} and Eq.~\eqref{eq : creation and annihilation op in momentum sp}, $\hat{V}$ is rewritten as
\begin{widetext}
\begin{align}
    \hat{V}
    =& 
    -\sum_{\boldsymbol{r}_1\boldsymbol{r}_2}
    V_{\sigma_1\sigma_2}(\boldsymbol{r}_1-\boldsymbol{r}_2)
    \hat{c}_{\boldsymbol{r}_1,\sigma}^\dagger \hat{c}_{\boldsymbol{r}_2,\sigma_2}
    \\
    =& 
    -\frac{1}{N_1N_2}
    \sum_{\boldsymbol{r}_1\boldsymbol{r}_2}
    \sum_{\boldsymbol{q}}
    \sum_{ \substack{\boldsymbol{k}_1 \in \mathrm{BZ}_1 \\ \boldsymbol{k}_2 \in \mathrm{BZ}_2}}
    v_{\sigma_1\sigma_2,\boldsymbol{q}}
    e^{-i\boldsymbol{r}_1\cdot(\boldsymbol{k}_1-\boldsymbol{q})}
    e^{i\boldsymbol{r}_2\cdot(-\boldsymbol{q}+\boldsymbol{k}_2)}
    \hat{c}^\dagger_{1,\boldsymbol{k}_1,\sigma_1}
    \hat{c}_{2,\boldsymbol{k}_2,\sigma_2}
    \\
    =& 
    -
    \sum_{\substack{\boldsymbol{G}_1 \in \mathcal{B}_1 \\ \boldsymbol{G}_2\in  \mathcal{B}_2 }}
    \sum_{\boldsymbol{q}}
    \sum_{ \substack{\boldsymbol{k}_1 \in \mathrm{BZ}_1 \\ \boldsymbol{k}_2 \in \mathrm{BZ}_2}}
    v_{\sigma_1\sigma_2,\boldsymbol{q}}
    e^{i(\boldsymbol{\tau}_1\cdot\boldsymbol{G}_1-\boldsymbol{\tau}_2\cdot\boldsymbol{G}_2)}
    \delta_{\boldsymbol{G}_1,-\boldsymbol{k}_1+\boldsymbol{q}}
    \delta_{-\boldsymbol{G}_2,-\boldsymbol{q}+\boldsymbol{k}_2}
    \hat{c}^\dagger_{1,\boldsymbol{k}_1,\sigma_1}
    \hat{c}_{2,\boldsymbol{k}_2,\sigma_2}
    \label{eq : before kronecker to dirac}
    \\
    =& 
    -
    \sum_{\substack{\boldsymbol{G}_1 \in \mathcal{B}_1 \\ \boldsymbol{G}_2\in  \mathcal{B}_2 }}
\sum_{ \substack{\boldsymbol{k}_1 \in \mathrm{BZ}_1 \\ \boldsymbol{k}_2 \in \mathrm{BZ}_2}}
v_{\sigma_1\sigma_2,\boldsymbol{k}_1+\boldsymbol{G}_1}
    e^{i(\boldsymbol{\tau}_1\cdot\boldsymbol{G}_1-\boldsymbol{\tau}_2\cdot\boldsymbol{G}_2)}
    \delta_{\boldsymbol{k}_1+\boldsymbol{G}_1,\boldsymbol{k}_2+\boldsymbol{G}_2}
    \hat{c}^\dagger_{1,\boldsymbol{k}_1,\sigma_1}
    \hat{c}_{2,\boldsymbol{k}_2,\sigma_2}.
\end{align}
Using the quasiperiodicity in Eq.~(\ref{eq : quasiperiodicity of c}), we arrive at the final expression in the commensurate system 
\begin{align}
    \hat{V}
    =& 
    -\sum_{\substack{\boldsymbol{G}_1 \in \mathcal{B}_1 \\ \boldsymbol{G}_2\in \mathcal{B}_2 }}
\sum_{ \substack{\boldsymbol{k}_1 \in \mathrm{BZ}_1 \\ \boldsymbol{k}_2 \in \mathrm{BZ}_2}}
    v_{\sigma_1\sigma_2,\boldsymbol{k}_1+\boldsymbol{G}_1}
     \delta_{\boldsymbol{k}_1+\boldsymbol{G}_1,\boldsymbol{k}_2+\boldsymbol{G}_2}
    \hat{c}^\dagger_{1,\boldsymbol{k}_1+\boldsymbol{G}_1,\sigma_1}
    \hat{c}_{2,\boldsymbol{k}_2+\boldsymbol{G}_2,\sigma_2}.
    \label{eq : final expression in the commensurate system}
\end{align}
\end{widetext}

Next, we move to the incommensurate system.
In the incommensurate limit $(S \nearrow \infty)$, we may approximate the $q$-summation with the $q$-integration.
From the definition of Kronecker's delta and Dirac's delta,  $f(q_0) = \sum_{q} \delta_{q,q_0} f(q)  = \int dq \delta(q-q_0) f(q)$, we may rewrite Eq.~\eqref{eq : before kronecker to dirac} as
\begin{align}
    &\sum_{\boldsymbol{q}}
    v_{\sigma_1\sigma_2,\boldsymbol{q}}
    \delta_{\boldsymbol{G}_1,-\boldsymbol{k}_1+\boldsymbol{q}}
    \delta_{-\boldsymbol{G}_2,-\boldsymbol{q}+\boldsymbol{k}_2}\nonumber
    \\
    =& 
    \int d^Dq
    v_{\sigma_1\sigma_2,\boldsymbol{q}}
    \delta(\boldsymbol{q}-\boldsymbol{k}_1-\boldsymbol{G}_1)
    \delta_{-\boldsymbol{G}_2,-\boldsymbol{q}+\boldsymbol{k}_2}
    \nonumber\\
    =& 
    v_{\sigma_1\sigma_2,\boldsymbol{k}_1+\boldsymbol{G}_1}
    \delta_{\boldsymbol{k}_1 + \boldsymbol{G}_1,\boldsymbol{k}_2 + \boldsymbol{G}_2},
\end{align}
with which we obtain the same form as Eq.~\eqref{eq : final expression in the commensurate system} in the incommensurate case as well. 
Accompanying to the approximation of $\sum_{\boldsymbol{q}}$ by $\int d^Dq$, we replace $\sum_{\boldsymbol{q}}$ in  Eq.~\eqref{eq : fourier V with alpha} with $S/(2\pi)^D\int d^Dq$.
This leads to
\begin{align}
    V_{\sigma_1\sigma_2}(\boldsymbol{r})
    =&
    \frac{\sqrt{S_1S_2}}{(2\pi)^D}
    \int d^Dq
    v_{\sigma_1\sigma_2,\boldsymbol{q}}e^{i\boldsymbol{q}\cdot\boldsymbol{r}},
    \\
    v_{\sigma_1\sigma_2,\boldsymbol{q}}
    =&
    \frac{1}{\sqrt{S_1S_2}}
    \int d^Dr V_{\sigma_1\sigma_2}(\boldsymbol{r})e^{-i\boldsymbol{q}\cdot\boldsymbol{r}},
    \\
    \hat{V}
    =& 
    -\sum_{\substack{\boldsymbol{G}_1 \in \mathcal{B}_1 \\ \boldsymbol{G}_2\in \mathcal{B}_2}}
\sum_{ \substack{\boldsymbol{k}_1 \in \mathrm{BZ}_1 \\ \boldsymbol{k}_2 \in \mathrm{BZ}_2}}
    v_{\sigma_1\sigma_2,\boldsymbol{k}_1+\boldsymbol{G}_1}
    \nonumber\\
    &\times \delta_{\boldsymbol{k}_1+\boldsymbol{G}_1,\boldsymbol{k}_2+\boldsymbol{G}_2}
    \hat{c}^\dagger_{1,\boldsymbol{k}_1+\boldsymbol{G}_1,\sigma_1}
    \hat{c}_{2,\boldsymbol{k}_2+\boldsymbol{G}_2,\sigma_2}.
\end{align}
These coincide with Eqs.~\eqref{eq : Fourier V1}, \eqref{eq : Fourier V2} and \eqref{eq : Interlayer coupling extended}.

\subsection{Decay of the interlayer coupling}\label{sec : decay of V}
The extended momentum space picture allows us to comprehend why an infinite number of foldings can be approximated by a finite number of foldings. In short, this is because $v_{\sigma\sigma^\prime,\boldsymbol{q}}$ decays rapidly against $|\boldsymbol{q}|$.
In Eq.~\eqref{eq : Interlayer coupling extended}, the amplitude $v_{\sigma_1\sigma_2}$ depends on the momentum in the extended momentum space.
Since the hopping amplitude from one site to another decays exponentially against the distance between two sites, we may expect $v_{\sigma\sigma^\prime,\boldsymbol{q}}$ also decays rapidly against $|\boldsymbol{q}|$ in the extended momentum space.
Indeed, proceeding researches \cite{PhysRevB.101.195425,Koshino_2015,PhysRevB.92.205108} model the interlayer coupling of two-dimensional layers as
\begin{align}
    V(\boldsymbol{r}) = V_0 e^{-\frac{\sqrt{r^2 + d^2}}{\xi}}.
\end{align}
Here, $d$ is the separation between two layers and $V_0$ and $\xi$ is a parameter to fit.
In the limit of $d =0$, we obtain $v_{\sigma\sigma^\prime,\boldsymbol{q}}$ analytically as
\begin{align}
    v_{\sigma\sigma^\prime,\boldsymbol{q}} = \frac{2\pi\xi^2V_0}{\sqrt{S_1S_2}} \frac{1}{(1+q^2\xi^2)^{\frac{3}{2}}}.
\end{align}
This function is a two-dimensional Lorentzian function and decays rapidly against $q$.  If we set a threshold for $v_{\sigma\sigma^\prime}$ and neglect the interlayer coupling below, momentum will also be restricted to the finite region in the extended momentum space. This leads to the finite number of foldings.

\begin{figure}[t]
    \centering
    \includegraphics[width=\linewidth]{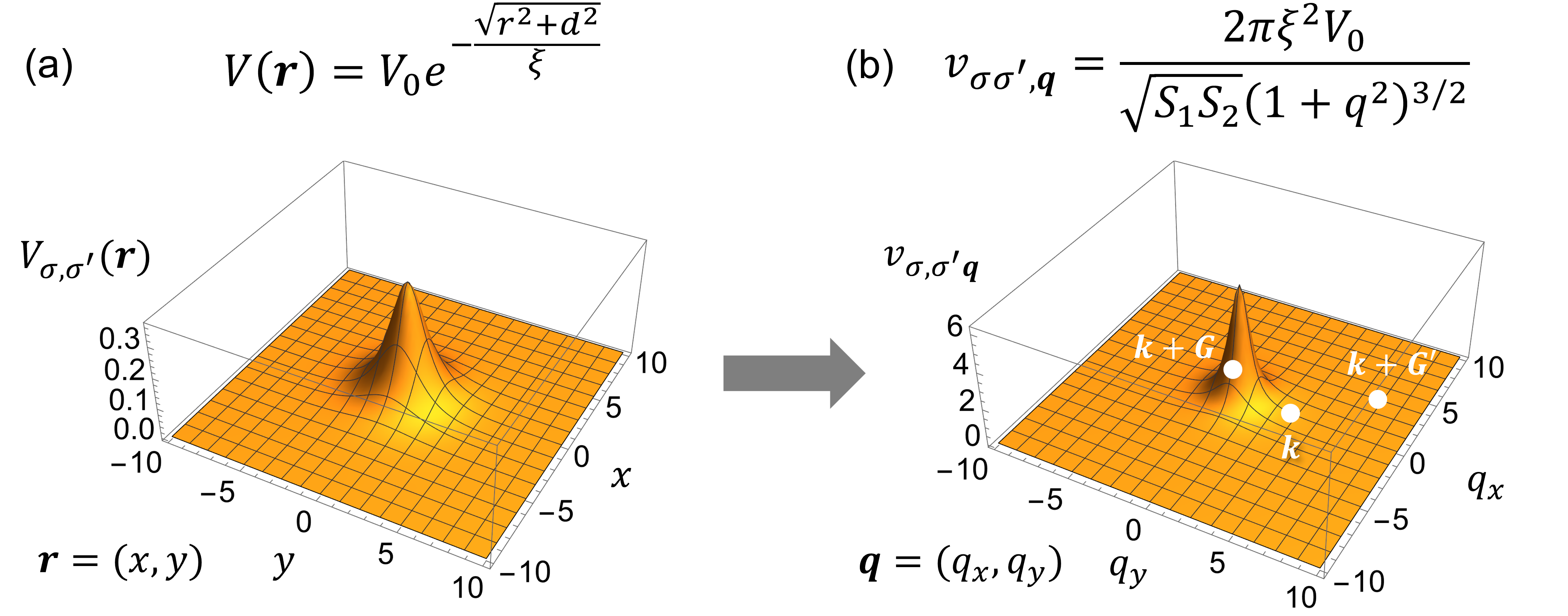}
    \caption{
    Interlayer coupling in the real space and momentum space.
    (a). Plot of $V_{\sigma\sigma^\prime}(\boldsymbol{r})$, where $\boldsymbol{r} = (x,y)$.
    (b). Plot of $v_{\sigma\sigma^\prime,\boldsymbol{q}}$, where $\boldsymbol{q} = (q_x,q_y)$.
    The interlayer layer coupling decays rapidly in both the real space and the momentum space. 
    We consider a momentum $\boldsymbol{k}$, and fold it to $\boldsymbol{k}+\boldsymbol{G}$ and $\boldsymbol{k}+\boldsymbol{G}^\prime$ using the reciprocal vectors $\boldsymbol{G}$ and $\boldsymbol{G}^\prime$.
    While $v_{\sigma\sigma^\prime,\boldsymbol{k}+\boldsymbol{G}}$ takes large value, $v_{\sigma\sigma^\prime,\boldsymbol{k}+\boldsymbol{G}^\prime}$ is almost zero. Therefore, we may drop $v_{\sigma\sigma^\prime,\boldsymbol{k}+\boldsymbol{G}^\prime}$ to reduce a small effective Hamiltonian.
    Here, we set $V_0, d, \xi$ and $S_lS_{l^\prime}$ to be $1$.
    }
    \label{fig:enter-label}
\end{figure}

\subsection{Integration over a large area}\label{sec : Integration over a large area}
\begin{figure} [t]
    \centering
    \includegraphics[width=\linewidth]{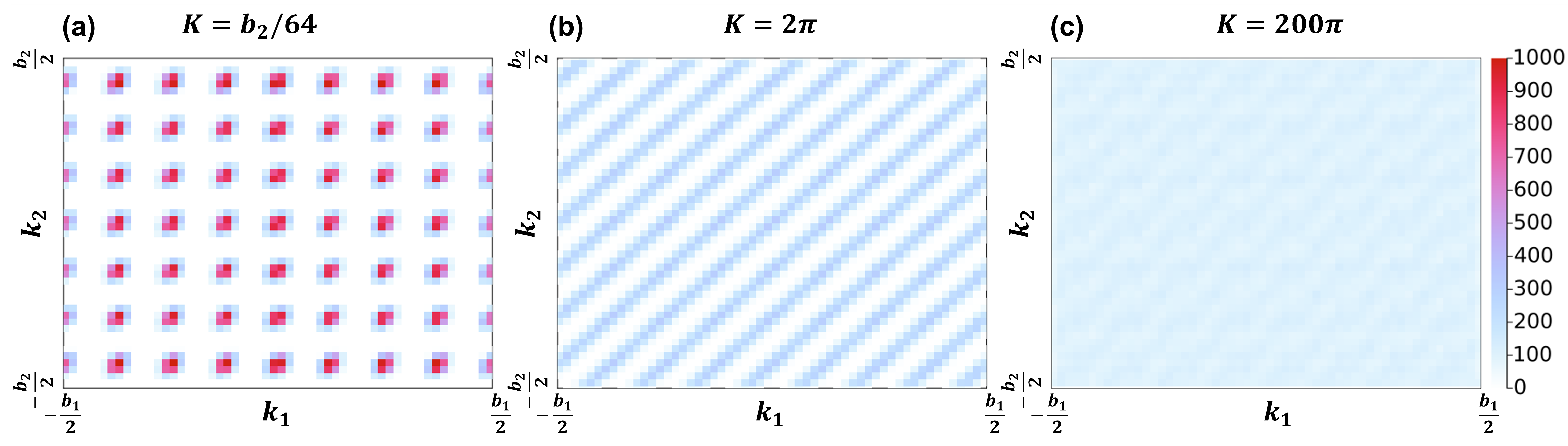}
    \caption{
    Two dimensional plot of the number of k-points mapped onto the torus.
    The system is the same with Sec.~\ref{sec : Convergence of error} with $F = 32$. 
    We sample k-point $\{k_1,\ldots, k_I\}$ from $-K$ to $K$ by $K/32$ and count the number of folded momenta $k_i+G_1+G_2 \ (k_i \in \{-K,-K+K/32,\ldots, K\}, G_l \in \boldsymbol{B}_l)$.
    Although the number of sampled points are the same, we can cover the torus homogeneously as we increase $K$.
    (a) When $K$ is small, blue regions are separated each other.
    (b) When $K = 2\pi$, which corresponds to the condition of Fig.~(\ref{fig : OP_incommensurate_real}). In this case, the torus is covered by the stripes.
    (c) When $K=200\pi$, blue region covers the torus densely.
    }
    \label{fig : DenseCoverofTorus}
\end{figure}

In numerical calculations, we replace the integral over the extended momentum space with the integral over a large area in the k-space and conduct the numerical integration.
While we are restricting the integration area from infinite area to finite area, this procedure is a good approximation due to the quasiperiodicity.
Here, we explain the validity of this approximation pictorially.
Let $f$ be a one-dimensional quasiperiodic function to integrate, and we approximate $\int dk f(k)$ with $\int_{-K}^{K} dk f(k)$. For visualization purposes, we assume $f$ is related to two-dimensional periodic function $F : (k_1,k_2) \mapsto F(k_1,k_2)$ defined on a torus $\mathrm{BZ}_1\times \mathrm{BZ}_2$ as $f(k) = F(k,k)$. 
Then $\int dk f(k) = \int dk F(k,k)$.
On the torus, we identify $k_1$ and $k_2$ as $(k_1,k_2) \sim (k_1+b_1,k_2) \sim (k_1,k_2+b_2)$. 
Since $f$ is quasi-periodic, while we move $k$ from $-\infty$ to $\infty$, $(k_1, k_2)$ also covers the torus in Fig.~\ref{fig : DenseCoverofTorus} densely.
Also in the numerical calculation, sampled k-points cover the torus densely. 
In Fig.~\ref{fig : DenseCoverofTorus}, we sample k-points as $\{-K,-K+K/32,\ldots,K\}$ and use the same system as in Sec.~\ref{sec : Convergence of error} with $F = 32$. 
While we increase $K$, blue regions change from separated islands [c.f. Fig.~\ref{fig : DenseCoverofTorus} (a)] to stripes [c.f. Fig.~\ref{fig : DenseCoverofTorus} (b)] and finally into a unified region [c.f. Fig.~\ref{fig : DenseCoverofTorus} (c)].
When $K = \infty$, blue points cover the torus homogeneously according to Weyl's equidistribution theorem \cite{stein2011fourier}.

\section{Intralayer interaction \label{sec : Intralayer interaction}}
In this section, we derive the expression for the interaction term in Eqs.~(\ref{eq : contact int qp}) and (\ref{eq : gap function finite momentum}).
We consider  the contact interaction on each layer,
where the interaction on layer $l$ is given by
\begin{align}
    H_{l,int} = -\sum_{\boldsymbol{r}_l} g_l 
    n_{l,\uparrow}(\boldsymbol{r}_{l}) 
    n_{l,\downarrow}(\boldsymbol{r}_{l}),
\end{align}
with $n_{l,\sigma}(\boldsymbol{r}_l) = \hat{c}_{l,\boldsymbol{r}_{l},\sigma}^\dagger \hat{c}_{l,\boldsymbol{r}_{l},\sigma}$.
Using the Fourier transformation 
\begin{align}
    n_{l,\sigma}(\boldsymbol{r}_l) &= \frac{1}{N_l}\sum_{\boldsymbol{k}} n_{l,\sigma}(\boldsymbol{k})e^{i\boldsymbol{k}\cdot\boldsymbol{r}_l}, \\
    n_{l,\sigma}(\boldsymbol{q})
    &= \sum_{\boldsymbol{k} \in \mathrm{BZ}_l}
    \hat{c}_{l,\boldsymbol{k}+\boldsymbol{q},\sigma}^\dagger \hat{c}_{l,\boldsymbol{k},\sigma},
\end{align}
we can rewrite the interaction term as
\begin{align}
    H_{l,int} 
    =& -\frac{g_l}{N_l^2}\sum_{\boldsymbol{r}_l}
    \sum_{\boldsymbol{k},\boldsymbol{k}^\prime \in \mathrm{BZ}_l}
    e^{-i(\boldsymbol{k}+\boldsymbol{k}^\prime)\cdot\boldsymbol{r}_l}
    n_{l,\uparrow}(\boldsymbol{k})
    n_{l,\downarrow}(\boldsymbol{k}^\prime), \nonumber
    \\
    =& -\frac{g_l}{N_l}
    \sum_{\boldsymbol{k} \in \mathrm{BZ}_l}
    n_{l,\uparrow}(\boldsymbol{k})
    n_{l,\downarrow}(-\boldsymbol{k}).
\end{align}
As a result, the second quantized form of the interaction term is given by
\begin{align}
    H_{l,int}
    =& -\frac{g_l}{N_l}
    \sum_{\boldsymbol{q}}
    \sum_{\boldsymbol{k}_1}
    \hat{c}_{l,\boldsymbol{k}_1+\boldsymbol{k},\uparrow}^\dagger 
    \hat{c}_{l,\boldsymbol{k}_1,\uparrow}
    \sum_{\boldsymbol{k}_2}
    \hat{c}_{l,\boldsymbol{k}_2-\boldsymbol{k},\downarrow}^\dagger \hat{c}_{l,\boldsymbol{k}_2,\downarrow},
    \nonumber
    \\
    =& -\frac{g_l}{N_l}
    \sum_{\boldsymbol{q}}
    \sum_{\boldsymbol{k}_1,\boldsymbol{k}_2}
    \hat{c}_{l,\boldsymbol{k}_1+\boldsymbol{q},\uparrow}^\dagger
    \hat{c}_{l,\boldsymbol{k}_2-\boldsymbol{q},\downarrow}^\dagger
    \hat{c}_{l,\boldsymbol{k}_2,\downarrow}
    \hat{c}_{l,\boldsymbol{k}_1,\uparrow}.
\end{align}
Next, we apply the mean-field approximation and define the SC gap function.
In single layer systems with the inversion and the time-reversal symmetry, the Cooper pairs takes nonzero value only when $\boldsymbol{k}_1+\boldsymbol{k}_2$ is zero in the BZ of layer $l$.
In extended BZ, this also means $\boldsymbol{k}_1+\boldsymbol{k}_2$ is a multiple of reciprocal vectors.
In heterolayer systems with inversion and time reversal symmetry, this means that the center of mass momentum of the Cooper pairs is the integer multiple of the reciprocal vectors.
Hence, fixing momentum $\boldsymbol{k}_1+\boldsymbol{k}_2$ to be a multiple of the reciprocal vectors, we obtain
\begin{widetext}
\begin{align}
    H_{l,int}
    =& -\frac{g_l}{N_l}
    \sum_{\boldsymbol{b}_M}
    \sum_{\boldsymbol{k},\boldsymbol{k}^\prime}
    \hat{c}_{l,\boldsymbol{k},\uparrow}^\dagger 
    \hat{c}_{l,\boldsymbol{b}_M-\boldsymbol{k},\downarrow}^\dagger
    \hat{c}_{l,\boldsymbol{b}_M-\boldsymbol{k}^\prime,\downarrow}
    \hat{c}_{l,\boldsymbol{k}^\prime,\uparrow}
    \nonumber
    \\
    =& 
    \sum_{\boldsymbol{b}_M}
    \left[
    \sum_{\boldsymbol{k}}
    \hat{c}_{l,\boldsymbol{k},\uparrow}^\dagger 
    \hat{c}_{l,\boldsymbol{b}_M-\boldsymbol{k},\downarrow}^\dagger
    \Delta_{l,\boldsymbol{b}_M}
    +
    \sum_{\boldsymbol{k}^\prime}
    \hat{c}_{l,\boldsymbol{b}_M-\boldsymbol{k}^\prime,\downarrow}
    \hat{c}_{l,\boldsymbol{k}^\prime,\uparrow}
    \Delta_{l,\boldsymbol{b}_M}^\dagger
    \right].
\end{align}
\end{widetext}
Here, the order parameter is defined as
\begin{align}
    \Delta_{l,\boldsymbol{b}_M} = 
    -\frac{g_l}{N_l}
    \sum_{\boldsymbol{k}^\prime}
    \langle
    \hat{c}_{l,\boldsymbol{b}_M-\boldsymbol{k}^\prime,\downarrow}
    \hat{c}_{l,\boldsymbol{k}^\prime,\uparrow}
    \rangle.
\end{align}

\section{Optical response}\label{sec : Optical response}
Here, we briefly explain the formula we used to calculate the optical absorption and shift current.

\subsection{Optical absorption}
Optical absorption $\alpha$ and optical susceptibility $\chi$ is related as
\begin{align}
    \alpha(\omega) = \frac{\omega\mathrm{Im}\chi(\omega)}{c}.
\end{align}
Here, $c$ is the speed of light.
Optical susceptibility $\chi(\omega)$ is expressed as
\begin{align}
    \chi(\omega) = \int \frac{d^Dk}{(2\pi)^D} \sum_{n,m} |r^a_{nm}|^2 f_{nm} \delta(\omega_{nm} - \omega) .
\end{align}
Here, $n,m$ is the label of eigenstates. $\omega_{nm} = E_n -E_m$ and $f_{nm} = f_n -f_m$ are the difference of energy $(E_n)$ and Fermi distribution function  $(f_n)$ for the band $n$ and $m$. 
$r^a$ is the position operator defined as
\begin{align}
    r^a_{nm} = \frac{v^a_{nm}}{i\omega_{nm}},
\end{align}
with the velocity operator $v^a_{nm} = \bra{n}\partial_{k_a} H(k) \ket{m}$. 
In the numerical calculations, we approximate the delta function with a Lorentzian
$
    \delta(x) = (\gamma/\pi)/(x^2 + \gamma^2)
$
with $\gamma = 1/16$.

\subsection{Shift current}
Shift current is the second order optical response, where a DC current $J_a$ is induced in the second order in the applied AC electric field $E_{b}$ as\cite{Cook2017}
\begin{align}
    J_{a} = \frac{c \epsilon_0 }{2}\kappa^{abb}(\omega) E_{b}(\omega)E_{b}(-\omega).
\end{align}
For normal conductors, $\kappa^{abb}$ is defined as \cite{Cook2017}
\begin{align}\label{eq : def kappa}
    \kappa^{abb}(\omega) = \frac{4\pi g_s e^3}{\hbar^3 \epsilon_0 c}\int \frac{d^D k}{(2\pi)^D} \sum_{n,m} f_{nm} I^{abb}_{nm}\delta(\omega_{nm} - \omega).
\end{align}
Here, $g_s = 2$ is the spin degeneracy, $\epsilon_0 $ is the vacuum permittivity, $e$ is the elementary charge. 
The integrand $I^{abb}$ is defined as
\begin{align}
    I^{abb}_{nm} = \mathrm{Re}[r^b_{mn}r^b_{nm;a}].
\end{align}
The covariant derivative of $r^a$ with respect to the direction $b$ is given by
\begin{widetext}
\begin{align*}
    r^a_{nm;b} = -\frac{1}{i\omega_{nm}}
    \left[
        \frac{v^a_{nm}\Delta^b_{nm} + v^b_{nm}\Delta^a_{nm}}{\omega_{nm}}
        -w^{ab}_{nm} 
        +\sum_{p \neq n,m} \left( \frac{v^a_{np}v^b_{pm}}{\omega_{pm}}
        -
        \frac{v^a_{np}v^b_{pm}}{\omega_{np}}
        \right)
    \right],
\end{align*}
\end{widetext}
with $w^{ab}_{nm} = \bra{n}\partial_{k_a}\partial_{k_b} H(k) \ket{m}$.
To apply the above expressions to the optical absorption and the shift current response in SCs, one only needs to redefine the velocity operator $v$ as \cite{Xu19}
\begin{align}
    v^{a}_{nm} =& \bra{n}\partial_{k_a} H_{BdG}(k) \tau_z\ket{m},
\end{align}
where $\tau_z$ denotes the particle-hole grading in SCs.

\bibliography{ref}

\end{document}